\documentclass[10pt,onecolumn]{aastex631}
\usepackage{graphicx}
\usepackage{amsmath}
\usepackage{amsfonts}
\usepackage{amssymb}%
\providecommand{\U}[1]{\protect\rule{.1in}{.1in}}

\usepackage[most]{tcolorbox}

\usepackage{tikz}
\usetikzlibrary{shapes,shadows,calc}
\usetikzlibrary{positioning,calc}

\usepackage{yfonts}
\usepackage[normalem]{ulem}

\usepackage[yyyymmdd,hhmmss]{datetime}

\newcommand{\beq}{\begin{equation}}
\newcommand{\eeq}{\end{equation}}
\newcommand{\ba}{\begin{array}}
\newcommand{\ea}{\end{array}}
\renewcommand{\L}{L_{52}}
\newcommand{\E}{E_{53}}

\newcommand{\M}{\dot{M}_{-6}}
\newcommand{\n}{n_{0,3}}

\newcommand{\ee}{\epsilon_{e,0}}

\newcommand{\D}{{\mathcal {D}}}
\newcommand{\R}{{\mathcal {R}}}
\renewcommand{\O}{{\mathcal {O}}}

\def\be{\begin{equation}}
\def\ee{\end{equation}}

\def\gsim{\raisebox{-0.3ex}{\mbox{$\stackrel{>}{_\sim} \,$}}}

\usepackage{color}
\usepackage{tabularx}
\usepackage{xcolor,colortbl}
\definecolor{LightCyan}{rgb}{0.88,1,1}
\definecolor{Chartreuse}{rgb}{0.5, 1, 0}

\newcommand{\I}{{\mathcal {I}}}
\newcommand{\J}{{\mathcal {J}}}

\newcommand{\chapter}[1]{}

\newcounter{chapnumber}
\usepackage{float}

\usepackage[dotinlabels]{titletoc}
\titlecontents{section} [7mm]
    {\addvspace{1mm}}
    {\contentslabel{7mm}}
    {}
    {\titlerule*[2.7mm]{.}\contentspage}
    [\addvspace{0mm}]

\shorttitle{GRBs emission constrained by Bethe-Heitler}
\shortauthors{B\'egu\'e, Samuelsson, and Pe'er}

\begin{document}

\title{
Bethe-Heitler signature in proton synchrotron models for gamma-ray bursts}
\author[0000-0003-4477-1846]{D. B\'egu\'e}
\affiliation{Bar Ilan University, Ramat Gan, Israel}
\correspondingauthor{D. B\'egu\'e : cayley38@gmail.com}

\author[0000-0001-7414-5884]{F. Samuelsson}
\affiliation{Department of Physics, KTH Royal Institute of Technology, \\
and The Oskar Klein Centre, SE-106 91 Stockholm, Sweden}

\author[0000-0001-8667-0889]{A. Pe'er}
\affiliation{Bar Ilan University, Ramat Gan, Israel}

\begin{abstract}
We study the effect of Bethe-Heitler (BeHe) pair production on a proton synchrotron model for the prompt
emission in gamma-ray bursts (GRBs). The possible parameter space of the model is constrained by consideration
of the synchrotron radiation from the secondary BeHe pairs. We find two regimes of interest. 1) At high bulk Lorentz
factor, large radius and low luminosity, proton synchrotron emission dominates and produces a spectrum in agreement
with observations. For part of this parameter space, a subdominant (in the MeV band) power-law is created by the
synchrotron emission of the BeHe pairs. This power-law extends up to few tens or hundreds of MeV. Such a signature
is a natural expectation in a proton synchrotron model, and it is seen in some GRBs, including GRB 190114C recently
observed by the MAGIC observatory.  2) At low bulk Lorentz factor, small radius and high luminosity, BeHe cooling
dominates. The spectrum achieves the shape of a single power-law with spectral index $\alpha = -3/2$ extending
across the entire GBM/Swift energy window, incompatible with observations. Our theoretical results can be used to further constrain the spectral analysis of GRBs in the guise of proton synchrotron models.
\end{abstract}


\section{Introduction}

The emission mechanism at the origin of the observed signal during the prompt phase of GRBs
remains unknown. Among the prime contenders are photospheric emission, released when the plasma becomes optically
thin \citep{Goo86,Pac86,MR00,DS02}, and synchrotron emission produced by relativistic particles accelerated by
shocks or magnetic reconnection once the flow is optically thin \citep{RM94,SNP96,DM98,ZY11}. In addition, protons may also contribute, either directly by synchrotron emission, or indirectly by emission from the secondaries produced in photo-hadronic and photo-pair processes \citep{AIM09,CK13,FPM21}.

When comparing models to spectral data, the most crucial difference between the aforementioned models are the
prediction for the low energy spectral slope $\alpha$, usually associated with the low-energy slope of the
Band model \citep{BMF93}. For photospheric emission models, the slope is expected to be around $\alpha = 0.4$
\citep{Bel10, PR11, BSV13, PL18}, unless the ejecta becomes transparent during the acceleration phase
\citep{Goo86, Pac86, BV14, RLA17}, in which case a steeper slope up to $\alpha = 1$ can be achieved. Slopes
shallower than $\alpha = 0.4$ can be obtained when considering geometrical effects such as emission from a
structured jet \citep{LPR13}, or subphotospheric dissipation \citep{PW05,Gia06,VB16}. Observationally, the footprint
of photospheric emission is seen in many GRB spectra, see \textit{e.g.} \cite{RP09, ARP20, DPR20}. Moreover, analysis
of GRB 090902B strongly supports a model where the emission is produced at the photosphere of a highly
relativistic outflow \citep{RAZ10,PZR12}. In the past years, photospheric models have been directly fitted to data
achieving good agreement \citep{ALN15, VGG17,SLR21}.   

Synchrotron models predict a low energy slope to be $\alpha = -2/3$
in the slow cooling regime and $\alpha = -3/2$ in the fast cooling regime. Slightly steeper slopes could be
obtained when considering the effects of inverse Compton cooling in the Klein-Nishina regime \citep{BDB09, NAS09,DBG11}.
Yet, most GRB spectra fitted with the Band function \citep{BMF93} are found incompatible with a synchrotron model:
this is known as the ''synchrotron line-of-death'' \citep{PBM98}. Recently, it was found that fitting a synchrotron model
directly to GRB spectra alleviates this problem\footnote{The spectral width was also proposed as a criteria to
further rule out synchrotron models \citep{AB15, YVG15}, but it was shown that the argument does not hold when
synchrotron models are fitted directly to the data \citep{Bur17}.} \citep{BBB19, ARP20}. The main reason for the disagreement
is that the Band function is a poor approximation of the synchrotron emission around the peak, giving poor
constraints when comparing the fitted results to the expectations from synchrotron models in a limiting energy window.
In addition, two independent analysis using \textit{Swift} X-ray \citep{ONG18} and optical
data \citep{ONG19} showed that the spectra of several GRBs require an additional break around
observed energy $\simeq 1$ keV, leading to the straightforward identification of the injection and cooling frequencies
of a synchrotron model. The spectral slopes below and between the breaks were also found compatible
with the expectation from synchrotron models with power-law distributed charged particles.

The closeness of the two identified breaks requires, within the framework of synchrotron models, that the emitting particles
be in the marginally fast cooling regime, with their cooling Lorentz factor $\gamma_{\rm c}$ nearly equal to their
injection Lorentz factor $\gamma_{\rm m}$. This requirement is difficult to account for if the radiating particles
are electrons \citep{BBG18}. Possible solutions include
the jet in jet model \citep{NK09,ZZ14,BBG18} or emission in a time dependent magnetic field \citep{UZ14b}.

Alternatively, it was proposed by \cite{GGO19} that protons could be the particles radiating synchrotron and
producing the main prompt MeV-peak. The observed requirement of marginally fast cooling is then
naturally fulfilled for emission radius in the order of
$10^{13}-10^{14}~$cm and bulk Lorentz factor of a few hundreds \citep{GGO19}, as expected for optically thin emission
models of GRBs \citep{RM94,DM98}. On the other hand, proton synchrotron models do not explain the observed spectral peak energy
clustering \citep{vMP20} and require a large magnetic luminosity $L_{\rm B} \gtrsim 10^{55}~$erg s$^{-1}$ \citep{FPM21}.

Synchrotron emission from protons and from the secondaries produced by photo-pion ( $p\gamma \rightarrow p + \pi^0$,
$p\gamma \rightarrow n+\pi^+$ and other channels producing two or more pions \citep{MER00,LLM07,HRS10}) and BeHe
($p\gamma\rightarrow p e^+ e^-$) interactions
was thoroughly studied in connection with ultra-high energy cosmic ray acceleration \citep{BD98, Tot98, RDF10},
high-energy PeV neutrino production \citep{Pet14} and high energy photon component observed by LAT and more recently by HESS and MAGIC \citep{GZ07,AIM09, CK13,SF20}. 
However, all those studies have in common the leptonic origin of the main MeV peak component, when it is not set to be a fiducial Band model.

\citet{FPM21} numerically studied a proton synchrotron model as the source of the main MeV peak similar to the one proposed by \cite{GGO19}. They concluded that emission from the secondaries produced
by either the BeHe-process or photo-pion interactions would be too bright to account for the optical constraints.
This is especially important since this result is inconsistent with
the claim that optical observations support synchrotron emission models \citep{ONG19}.
However, the bursts used by \cite{ONG19} and afterwards by \cite{FPM21} are very long duration bursts ($T_{90} > 70-80$s),
which is required to have simultaneous optical observations. Thus, this small subset of bursts is not necessarily
representative of the full GRB population, nor of the emission mechanism producing the early episodes of a GRB.
Indeed, it was suggested that the emission mechanism might change throughout the burst episodes
(\textit{e.g.} \cite{ZZC18, Li19}).
It is therefore interesting to find predictions of proton synchrotron models that do not rely on optical data,
to be able to test the model on shorter duration bursts.

In this paper, we assume that the prompt emission is due to proton synchrotron and derive constraints on the model.
We present analytical estimates of the effect of BeHe pair production and the pairs subsequent radiation. We
identify two emission regimes: 1) proton synchrotron dominated emission regime with little contribution from other processes,
2) BeHe pair dominated emission regime, leading to a spectrum incompatible with observations. We further describe the transition between these two extreme regimes in which a subdominant power-law from the BeHe pair synchrotron radiation appears across the MeV band, as observed in some GRBs (\textit{e.g.} \cite{VGG17,CPB20}).

The paper is organized as follows. In Section
\ref{sec:2}, we identify the parameter space for each of the three regimes mentioned above by comparing the timescales
of synchrotron emission to that of BeHe pair production. Section \ref{sec:3} details the modification to the spectrum
due to synchrotron radiation from the pairs produced by the BeHe process. Discussion with an emphasis on GRB 190114C is
given in Section \ref{sec:4}.

\section{Constraining proton synchrotron emission models by Bethe-Heitler cooling}
\label{sec:2}

In this section, we compare the timescale of proton synchrotron emission to that of BeHe pair production. If the protons
cool too quickly from BeHe pair creation, the secondary emission from the pairs can greatly affect the observed spectrum.
The valid parameter space for proton synchrotron models can thus be constrained.

Consider an emission region expanding relativistically with Lorentz factor $\Gamma$, emitting radiation at a distance
$r$ from a central engine, and threaded by a magnetic field of comoving strength $B$. The comoving dynamical
time is given by $t_{\rm dyn} = r/(\Gamma c)$, where $c$ is the speed of light. In a marginally fast cooling scenario,
relativistic particles, here protons, are assumed to be steadily injected into a power-law with index $-p$ above some
injection Lorentz factor $\gamma_{\rm p,m}$. We present our results for $p = 2.5$ and $p = 3.5$. On the one hand, the
value of $p \sim 2.5$ is expected in many dissipation and acceleration scenarios (\textit{e.g.} \citet{BO98, KGG00}),
albeit softer values can also be obtained from simulations \citep{SSA13, CCM19,CSS20}. On the other hand, synchrotron
fits to the GRB spectra require an average value of $p = 3.5$ \citep{BBB19}. Marginally fast cooling implies that
$\gamma_{\rm p,m} \sim \gamma_{\rm p,c}$, where $\gamma_{\rm p,c}$ is the characteristics proton cooling Lorentz factor.
We write $\gamma_{\rm p,m} = \xi \gamma_{\rm p,c}$. In this paper, we assume $\xi \gtrsim 1$, i.e., the protons are
fast cooling albeit marginally. This implies that protons efficiently radiate most of their energy, while satisfying
the observed spectral constraints.

The comoving cooling time for protons with Lorentz factor $\gamma_{\rm p}$ emitting synchrotron radiation is
\begin{align}
t_{\rm synch} = \frac{6\pi c m_{\rm p}}{B^2 \gamma_{\rm p} \left (\frac{m_{\rm e}}{m_{\rm p}} \right)^2 \sigma_T},
\end{align} 
and the frequency of the synchrotron spectral peak is   $
\nu_{\rm peak} = (4/3) \Gamma q \gamma_{\rm p,m}^2 B /(\pi c m_{\rm p})$,
where we used $\nu_{\rm peak}$ as the frequency without redshift correction, \textit{i.e.} in the frame of the burst.
The observed frequency is $\nu^{\rm obs} = \nu_{\rm peak}/(1+z)$. In those equations, $m_{\rm p}$ and $m_{\rm e}$ are the proton and
electron masses, $q$ is the elementary charge and $\sigma_{\rm T}$ is the Thompson cross section.

Setting the dynamical timescale and the cooling timescale equal, $t_{\rm dyn} \sim t_{\rm synch}$, gives the magnetic
field and the proton cooling Lorentz factor
\begin{align}
B &= \frac{2 \sqrt[3]{6\pi q} c m_{\rm p}^{\frac{5}{3}}}{ m_{\rm e}^{\frac{4}{3}}  \sigma_{\rm T}^{\frac{2}{3}}} \frac{\Gamma \xi^{\frac{2}{3}} }{r^{\frac{2}{3}}\sqrt[3]{\nu_{\rm peak}}} \sim  3.3 \times 10^6 ~{\rm G}  ~\frac{\Gamma_2 \xi^{\frac{2}{3}} }{r_{14}^{\frac{2}{3}} \nu_{\rm MeV}^{\frac{1}{3}}},  \\
\gamma_{\rm p,c} &= \frac{\sqrt[3]{6\pi m_{\rm e}^2 \sigma_{\rm T}}}{4 \sqrt[3]{m_{\rm p} q^2}} \frac{\sqrt[3]{r \nu_{\rm peak}^2}}{\Gamma \xi^{\frac{4}{3}}} \sim 1.4 \times 10^4 ~ \frac{r_{14}^{\frac{1}{3}} \nu_{\rm MeV}^{\frac{2}{3}}}{\Gamma_2 \xi^{\frac{4}{3}}},
\end{align}
where we have replaced $\gamma_{\rm p,m}$ by $\xi \gamma_{\rm p,c}$ and have used the notation
$Q_x = Q/10^x$. Here, $\nu_{\rm MeV} = h\nu_{\rm peak}/1~$MeV is the peak energy from the proton
synchrotron, normalised to the value 1 MeV in agreement with observations, and
$h$ is Planck's constant.

The comoving photon peak energy is $h\nu_{\rm m} = h\nu_{\rm peak}/(2\Gamma) = 5.0 ~{ \rm  keV  } ~\nu_{\rm MeV} \Gamma_2^{-1}$.
Most of the accelerated protons\footnote{Our model only describes the non-thermal population of protons. Colder protons
are also present in the flow, but we discard their contribution to the overall emission process.} have Lorentz factor
$\gamma_{\rm p,c}$. Therefore, for the bulk number of accelerated protons interacting with photons at the peak,
one gets
\begin{align}
\gamma_{\rm p,c} \, h\nu_{\rm m} = 67.5 ~{ \rm MeV} ~ \frac{r_{14}^{\frac{1}{3}} \nu_{\rm MeV}^{\frac{5}{3}}}{\Gamma_2^2 \xi^{\frac{4}{3}}}, \label{eq:prod_gps_nup}
\end{align}
which satisfies the threshold requirement for the BeHe-process ($\gamma_{\rm p,c} \, h\nu_{\rm m} > 2 m_ec^2$) unless $r_{14}^{\frac{1}{3}} \nu_{\rm MeV}^{\frac{5}{3}} < 0.015 \times \Gamma_2^2 \xi^{\frac{4}{3}}$. We note
that the lowest energy protons cannot satisfy the energy threshold for photo-pion interaction
$\gamma_{\rm p,c} \, h \nu_{\rm m}^{'} < 135~ {\rm MeV}$, and it is therefore expected that neutrino production in this
model be small. We comment further on the relevant cooling times in the discussion section. 
The model can be further constrained by the tight constraints from the IceCube \citep{AAA17} and Antares \citep{AAA17_Antares} experiments, as shown by \cite{FPM21} and \cite{PTP21}.

Having verified that all accelerated protons are energetic enough to satisfy the threshold
of BeHe pair creation, we now estimate the cooling of proton by the BeHe. This timescale is a function
of the comoving photon spectrum near the peak of their distribution, which itself depends on the comoving proton density\footnote{In principle, the analysis
can be done without computing the proton density, as the photon density can be directly expressed in terms of luminosity,
radius, Lorentz factor and peak energy. This does not change the dependency on the parameters. Here, we chose to make
explicit use of the proton density. }. Let $L^{\rm obs}$ be the observed isotropic photon luminosity of the burst.
Assuming the main emission mechanism is proton synchrotron, the number of radiating protons $N_{\rm p}$ is
\begin{align}
N_{\rm p} \sim \frac{L^{\rm obs}}{P_{\rm synch}^{\rm obs} (\gamma_{\rm p,m})} = \frac{L^{\rm obs}}{\frac{4}{3} c \gamma_{\rm p,m}^2 \Gamma^2 u_{\rm B} \sigma_{\rm T} \left ( \frac{m_{\rm e}}{m_{\rm p}}\right )^2} \sim 1.64 \times 10^{48} ~ \frac{L_{52} r_{14}^{\frac{2}{3}}}{\Gamma_2^2 \nu_{\rm MeV}^{\frac{2}{3}} \xi^{\frac{2}{3}}},
\end{align}
where $P_{\rm synch}^{\rm obs} (\gamma_{\rm p,m})$ is the observed synchrotron power emitted by a single proton with Lorentz
factor $\gamma_{\rm p,m}$ and $u_{\rm B} = {B}^2/(8\pi)$ is the comoving magnetic energy density. 

For the comoving volume, we use $V = 4 \pi r^2 (r/\Gamma)$ (\textit{e.g.} \cite{Pee15}), and therefore, the comoving density
of radiating protons is given by
\begin{align}
n_{\rm p} \sim \frac{N_{\rm p}}{V} = & 1.3 \times 10^{7} ~ {\rm cm}^{-3} ~ \frac{L_{52}}{\Gamma_2 r_{14}^{\frac{7}{3}} \nu_{\rm MeV}^{\frac{2}{3}} \xi^{\frac{2}{3}}}, \label{eq:proton_density}
\end{align}
To normalise the photon spectrum, it is assumed that the whole power radiated by protons with Lorentz factor $\gamma_{\rm p,m}$
is emitted at $\nu_{\rm m}$. Thus, the peak spectral energy density is
\begin{align}
u_{\nu_m} \sim \frac{n_{\rm p} P_{\rm synch}(\gamma_{\rm p,m})}{\nu_{\rm m}} t_{\rm dyn} =  2.2 \times 10^{-10} ~ {\rm  erg}~ {\rm cm^{-3}}~ {\rm Hz^{-1}}~L_{52} \Gamma_2^{-1} r_{14}^{-2} \nu_{\rm MeV}^{-1}, \label{eq:spectral_energy_density_synchrotron_proton}
\end{align} 
where $P_{\rm synch} = \Gamma^{-2} P_{\rm synch}^{\rm obs}$. Since protons with Lorentz factor $\gamma_{\rm p, m}$
mostly interact via BeHe process with photons close to the peak, only the shape of the photon spectrum around the
peak affects the cooling rate by the BeHe process. The photon distribution close to the peak is well
approximated by the synchrotron radiation of the protons even when $t_{\rm BeHe}\sim t_{\rm synch}$,
where $t_{\rm BeHe}$ is the BeHe cooling timescale.
This can be understood because the photons produced by the BeHe pairs are at different energies (see section \ref{sec:3}),
where the cross-section is smaller. Furthermore, when $t_{\rm BeHe}\sim t_{\rm synch}$, the proton distribution
function is not strongly changed below $\gamma_{\rm p, m}$.
Therefore, in the marginally fast cooling scenario considered in this paper $(\xi \geq 1)$, the comoving
photon spectrum around the peak is obtained as
\citep{SPN98}
\begin{align}
n_\nu =& \frac{1}{h \nu} u_{\nu_{\rm m}} \left \{
\begin{aligned}
& \left ( \frac{\nu}{\nu_{\rm c}} \right )^{\frac{1}{3}} \left ( \frac{\nu_{\rm c}}{\nu_{\rm m}} \right )^{-\frac{1}{2}} & & \nu < \nu_{\rm c},\\
& \left ( \frac{\nu}{\nu_{\rm m}} \right )^{-\frac{1}{2}} & & \nu_{\rm c} < \nu < \nu_{\rm m}, \\
& \left ( \frac{\nu}{\nu_{\rm m}} \right )^{-\frac{p}{2}} & & \nu_{\rm m} <  \nu .
\end{aligned} \right.  \label{eq:synch_spectrum_fast_cooling}
\end{align}
where $p$ is the index of the proton spectrum.

In Appendix \ref{app:appendix_BeHe_cooling_rate_simplification}, we obtain the cooling rate of protons by BeHe pair production ($p\gamma\rightarrow p e^+ e^-$)
following the prescription of \citet{CZS92}. The cooling is a function of photon energy in the proton rest frame, thus,
one has to integrate the photon distribution over energy and angle. This is done when generating the figures, which therefore show exact results in the case of an isotropic photon distribution. Here, we present approximate analytical estimates to demonstrate how the cooling varies with the parameters. Using Equations \eqref{app:eq_dgammadt_BeHe}
and \eqref{app:eq_BeHe_cooling}, and assuming  $\xi =1$ for simplicity, the BeHe timescale is given by
\begin{align}
t_{\rm BeHe}^{-1} = -\left. \frac{1}{\gamma_{\rm p}} \frac{d\gamma_{\rm p}}{dt} \right |_{\rm BeHe} = \left \{ \begin{aligned}
&  8.3 \times 10^{-2}  ~ {\rm s^{-1}} ~ \Gamma_2^{-\frac{2}{3}} L_{52} r_{14}^{-2} \gamma_{\rm p}^{-\frac{1}{3}} \nu_{\rm MeV}^{-\frac{4}{3}} & ~~~~~ & 2 x_{\rm m} \gamma_{\rm p} > \kappa_0,  \\
&  4.8 \times 10^{-7} ~ {\rm s^{-1}} ~   L_{52} \gamma_{\rm p}^{\frac{5}{4}} \nu_{\rm MeV}^{\frac{1}{4}} r_{14}^{-2} \Gamma_2^{-\frac{9}{4}}  & ~~~~~ & {\rm else,}
\end{aligned} \right.
\label{eq:t_BeHe}
\end{align}
where $x_{\rm m} = h \nu_{\rm m}/(m_e c^2)$ and $\kappa_0 = 40$ is found to provide an adequate approximation
for the cooling rate, see Appendix \ref{app:appendix_BeHe_cooling_rate_simplification}. The parameters $\kappa_0$ is introduced to simplify the expression for the BeHe cooling. It is roughly the photon energy (in the proton rest frame) corresponding to the maximum cooling rate by the BeHe process.  To obtain the numerical
value in the bottom expression, $p$ was set to 2.5. The expression and numerical value for a different value of
$p$ can be obtained by using
Equations \eqref{app:eq_dgammadt_BeHe} and \eqref{app:eq_BeHe_cooling}. The first line of Equation \eqref{eq:t_BeHe}
is for protons that mostly interact with photons below $\nu_{\rm m}$, while the second line describes
protons interacting with photons with frequency higher than $\nu_{\rm m}$. An estimate of the ratio between
the BeHe cooling time and the synchrotron cooling time at the injection Lorentz factor $\gamma_{\rm p,m}$ is given by
\begin{align}
\left. \frac{t_{\rm BeHe}}{t_{\rm synch}}\right |_{\gamma_{\rm p,m}} = \left \{
\begin{aligned}
& 8.57  ~ r_{14}^{\frac{10}{9}} \nu_{\rm MeV}^{\frac{14}{9}} \Gamma_2^{\frac{4}{3}} L_{52}^{-1} & ~~~~~ & 2 x_{\rm m} \gamma_{ \rm  p, m} > \kappa_0,  \\
& 4.3 \times 10^{-1} ~  \Gamma_2^{\frac{9}{2}} r_{14}^{\frac{7}{12}} \nu_{\rm MeV}^{-\frac{13}{12}} L_{52}^{-1} & ~~~~~ & {\rm else,}
\end{aligned}
\right.
\label{eq:ratio_cooling_at_gamma_m}
\end{align}
where we used the fact that $t_{\rm synch} = t_{\rm dyn}$ for $\gamma_{\rm p,m}$. This results therefore indicates
similar timescale for fiducial parameters. As noted above, the similarities of the time scales implies that the peak of the proton synchrotron is not substantially modified by the BeHe process.



\begin{figure}
\centering
\includegraphics[width=0.99\textwidth]{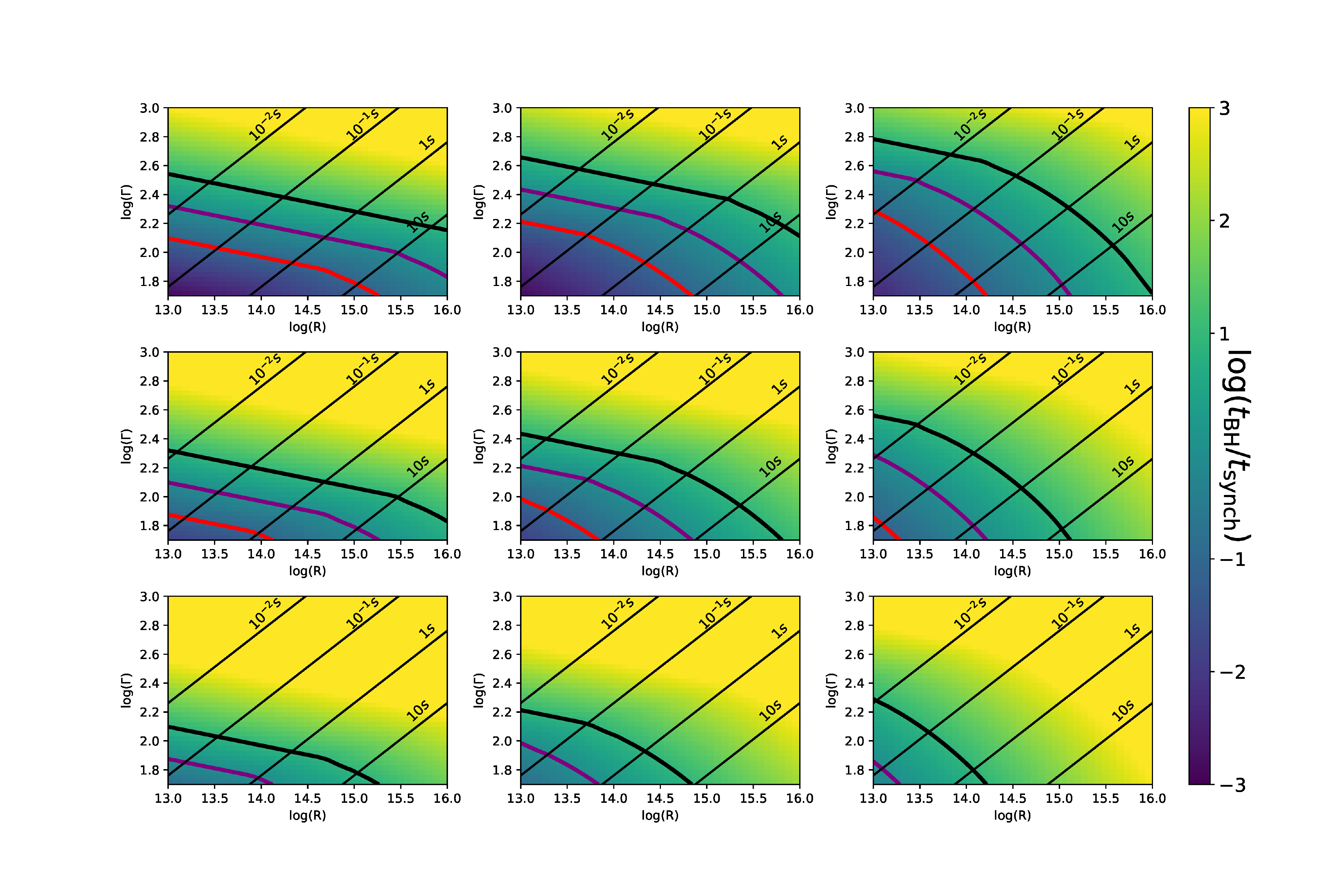}
\caption{Comparison of cooling rate between the BeHe and synchrotron processes at $\gamma_{\rm p,m}$. From top to bottom the emitted luminosity is $10^{54}$, $10^{53}$ and $10^{52}$ erg s$^{-1}$, while from left to right the observed spectral peak frequency is $100$ keV, $300$ keV and 1 MeV. The purple, red and black thick lines correspond to $t_{\rm BeHe} = t_{\rm synch}$, $t_{\rm BeHe} = 0.1 t_{\rm synch}$, and $t_{\rm BeHe} = 10 t_{\rm synch}$ respectively. The thin black lines show the variability time expected from the Lorentz factor and radius $t_{\rm var} \sim r/(\Gamma^2 c)$ for selected variability time $10^{-2}$s, $10^{-1}$s, $1$s and $10$s. The figure is made with $p = 2.5$. A value of $p = 3.5$ slightly increases the valid parameter space for proton synchrotron models by increasing the timescale ratio $t_{\rm BeHe}/t_{\rm synch}$ at small radii. }
\label{fig:cooling_comparison}
\end{figure}

Figure \ref{fig:cooling_comparison} shows the ratio of cooling times at $\gamma_{\rm p,m}$ assuming $\xi = 1$ for
different parameter choices. It is obtained by direct integration of Equation \eqref{app_eq:dgp_dt}. In computing
this figure, we have assumed that the proton distribution function is only modified by synchrotron losses. This
assumption breaks when BeHe cooling strongly dominates in the lowest domain of each panel in Figure
\ref{fig:cooling_comparison}. From top to bottom, the emitted luminosity is $10^{54}$, $10^{53}$, and $10^{52}$
erg s$^{-1}$ and from left to right the observed spectral peak energy is $100$ keV, $300$ keV, and 1 MeV,
respectively. Figure \ref{fig:cooling_comparison} is made with $p = 2.5$. A softer value of $p$ increases
the timescale ratio for the high-energy branch, \textit{i.e.}, it only affects the left-most part in the
panels in Figure \ref{fig:cooling_comparison}. For $p = 3.5$ as compared to $2.5$, the timescale ratio
increases by a factor $\sim 2$ for an order of magnitude decrease in radius.

From this figure, as well as from Equation
\eqref{eq:ratio_cooling_at_gamma_m}, one can identify two extreme regimes. For low luminosities $L^{\rm obs}$, high
Lorentz factor $\Gamma$ and large radius $r$, the protons are largely unaffected by BeHe pair creation (yellow
region in Figure \ref{fig:cooling_comparison}).
In this scenario, the observed spectrum is due to the synchrotron emission from the marginally fast
cooling protons as described in \citet{GGO19}, without any modification by BeHe. This shows
that explaining GRB prompt spectra with proton synchrotron requires high bulk Lorentz factor $\Gamma \gtrsim 300$,
in agreement with the analysis of \cite{FPM21} who used optical constraints. On the other end, when
$\Gamma$ and $r$ are small and/or $L$ is high, the BeHe process dominates the cooling
(dark region in Figure \ref{fig:cooling_comparison}). 
In this regime, the synchrotron photons from the very fast cooling pairs quickly outnumber the
proton synchrotron photons, leading to even more rapid BeHe pair creation. Therefore, most of the available
proton energy is extracted by the BeHe pairs. The cooling Lorentz factor of the pairs is
$\gamma_{\pm, {\rm c}} \sim 1$, corresponding to a cooling break in the observed spectrum at $\sim 10~$eV, whereas
the $\nu F_\nu$-peak energy associated to the synchrotron from the created BeHe pairs is at $\sim 100~$MeV (see Equation \eqref{eq:ratio_peak_energy}). Thus, the
observed spectrum consists of a single power-law with $F_\nu \propto \nu^{-1/2}$ between these two
energies, clearly incompatible with observed GRB spectra. In between the two extreme regimes when
the cooling timescales are comparable, signatures from both processes can be seen in the spectrum,
and we explore this scenario in Section \ref{sec:3}.


\section{Spectral signature of Bethe-Heitler pairs}
\label{sec:3}

In this section, we obtain predictions for the comoving pair distribution and their emission spectrum in the case where the cooling via synchrotron and BeHe are comparable. In this situation, the proton distribution at $\gamma_{\rm p,m}$ is only marginally affected by BeHe cooling. This implies that the photon spectrum at the peak energy around 1 MeV (which is the optimal photon energy for BeHe pair creation; see Equation \eqref{eq:prod_gps_nup} and Appendix \ref{app:appendix_BeHe_cooling_rate_simplification}) is not strongly modified by synchrotron radiation from the secondaries. If the secondary emission from the pairs do substantially contribute to the BeHe cooling of the protons, the BeHe pair creation becomes exponential in time and we are instead in the regime where BeHe dominates. Here, we use the proton synchrotron photons as targets to compute the rate at which pairs are created, namely in the case $t_{\rm BeHe} \gsim t_{\rm synch}$.



A proton with Lorentz factor $\gamma_{\rm p}$ produces electrons and positrons with typical Lorentz factor
$\gamma_\pm = \kappa_{\rm e} (m_{\rm p}/m_{\rm e})\gamma_p$, where $\kappa_{\rm e}$ is the inelasticity. The
dependence of the inelasticity on $\gamma_{\rm p} x$, where $x$ is the target photon energy in units of
electron rest mass, can be found in \cite{MPK05}. For photons at the peak energy, $\gamma_{\rm p} x$ is
given by Equation \eqref{eq:prod_gps_nup} and is of the order of a few to a few hundreds. Looking at Figure 1
of \cite{MPK05} for those values of $\gamma_{\rm p} x$, the inelasticity is found to vary
between $10^{-3}$ and $10^{-4}$, giving an average pair Lorentz factor between $\gamma_\pm \sim 2\gamma_{\rm p}$
and $\gamma_\pm \sim \gamma_{\rm p}/5$. Considering that most of the protons have Lorentz factor
$\gamma_{\rm p,m} \sim \gamma_{\rm p,c}$, we consider that all pairs are created with Lorentz factor
$\gamma_{\pm,{\rm m}} \equiv \kappa_{\rm e} (m_{\rm p}/m_{\rm e})\gamma_{\rm p,m}$. In other words, we
neglect the contribution of higher energy protons in the creation of pairs with higher energies. This
effect only changes the very high energy photon spectrum, which is likely to be absorbed by pair creation.
In addition, since the pairs are fast cooling, which follows as the protons are marginally fast cooling,
they obtain a Lorentz factor much smaller than their initial Lorentz factor in one dynamical timescale
$\gamma_{\pm,c} \ll \gamma_{\pm,{\rm m}}$, and therefore the exact details of their injection is lost via
their cooling.

The pair production rate by BeHe is given by \cite{CZS92}, but cannot be analytically integrated for a general
photon spectrum. Analytical estimates in some specific cases were provided by \cite{PM15b}.
We provide the integral expression used in our numerical computation in Appendix \ref{app:appendix_pair_production_rate_simplification}. 
In order to get analytical estimates of the number of pairs, we write the ratio between the synchrotron power
and BeHe power to be equal to the ratio of their timescales:
\begin{align}
    \frac{P_{\rm BeHe}^{\rm tot}}{N_{\rm p} P_{\rm synch}} = \left (\frac{t_{\rm BeHe}}{t_{\rm synch}} \right )^{-1},
\label{eq:power_ratio}
\end{align}
where $P_{\rm BeHe}^{\rm tot}$ it the power emitted by all BeHe pairs. Using $P_{\rm BeHe}^{\rm tot} = \gamma_{\pm,{\rm m}} m_{\rm e} c^2 \dot n_\pm V$, which is valid since the pairs are fast cooled, one obtains
\begin{align}
    \dot n_\pm = n_{\rm p} 
    \left (\frac{t_{\rm BeHe}}{t_{\rm synch}} \right )^{-1} \frac{P_{\rm synch}}{\gamma_{\pm,{\rm m}} m_{\rm e} c^2} = \left \{ 
    \begin{aligned}
    & 4.6 \times 10^8 ~ {\rm cm }^{-3} {\rm s}^{-1} ~ L_{52}^{2} \Gamma_2^{-\frac{4}{3}} r_{14}^{-\frac{40}{9}} \kappa_{\rm e, -4}^{-1} \nu_{\rm MeV}^{-\frac{20}{9}}  & ~~~~~ & 2 x_{\rm m} \gamma_{\rm p, m} > \kappa_0, \\
    & 9.2 \times 10^9 ~ {\rm cm }^{-3} {\rm s}^{-1} ~ L_{52}^{2} \nu_{\rm MeV}^{\frac{5}{12}} \Gamma_2^{-\frac{9}{2}} r_{14}^{-\frac{47}{12}} \kappa_{\rm e, -4}^{-1} & ~~~~~ & {\rm else.}
    \end{aligned}\right.
\label{eq:n_pm_dot}
\end{align}
We note the very strong dependence on the parameters, specifically the radius and the peak energy. This means
that in principle both scenarios with high and low pair yield are possible.

Assuming that all pairs are produced at Lorentz factor $\gamma_{\pm,{\rm m}}$ and that pair annihilation
is negligible (see Section \ref{sec:4}), the continuity equation for
the pairs can be solved to obtain the pair distribution. This computation is done
in Appendix \ref{sec:appendix_pair_spectrum}, and gives
\begin{align}
n_\pm (\gamma_{\rm e}) = \frac{m_{\rm e} c^2 \dot n_\pm}{P_{\rm e}(\gamma_{\pm,{\rm m}})} \left( \frac{\gamma_{\pm,{\rm m}}}{\gamma_{\rm e}} \right )^{2} H(\gamma_{\pm,{\rm m}} - \gamma_{\rm e}), \label{eq:pair_production_rate}
\end{align}
namely a single power-law with index $-2$ extending from $\gamma_{\pm,{\rm m}}$ down to $\gamma \sim 1$.
Here, $P_{\rm e}$ is the synchrotron power emitted by an electron and $H$ is the Heaviside function. We note that
at low energies the pair distribution should substantially deviate from this power-law because of strong
synchrotron self-absorption heating and pair annihilation. Our analysis also neglects the electrons originally present in the flow (see the discussion in Section \ref{sec:4}).

We now estimate the emerging spectrum. The emitted synchrotron spectrum from the pair distribution
in Equation \eqref{eq:pair_production_rate} is a fast cooling power-law with $F_\nu \propto \nu^{-1/2}$.
It extends from an observed peak frequency of
\begin{align}
\nu_{ {\rm m}, \pm} = \left(\frac{\gamma_{\pm,{\rm m}}}{\gamma_{\rm p,m}}\right)^{2} \left(\frac{m_{\rm p}}{m_{\rm e}}\right) \nu^{\rm obs} = 61.9 ~{\rm MeV}~ \kappa_{\rm e,-4}^2 \nu_{\rm MeV} \label{eq:ratio_peak_energy}
\end{align}
down to sub-keV energies. The synchrotron frequency of the pairs linearly depends on the observed peak frequency.
It also indirectly depends on the other model parameters via the value of the inelasticity. Larger values of
$\kappa_{\rm e}$ results from smaller values of $\gamma_{\rm p,m} \nu_{\rm m}$, i.e., larger Lorentz factor
and peak frequency, and/or smaller radius and luminosity, see Equation \eqref{eq:prod_gps_nup}. The shape of
the spectrum above $\nu_{\pm,{\rm m}}$ depends on the shape of the proton distribution function and of the pair
injection details. Since we are only giving analytical estimates, it is out of the scope of this paper to
account for a detailed analysis at those energies. We note however that emission at GeV and eventually TeV
energies are constrained by the LAT instrument on-board Fermi (\textit{e.g.} \cite{GPW11}). The normalization
of the spectrum is either obtained from the pair distribution function in Equation \eqref{eq:pair_production_rate},
or by considering the ratio of the synchrotron power to the BeHe power in Equation \eqref{eq:power_ratio}.
Indeed, the value and parameter dependence of the ratio between
the proton synchrotron peak and the BeHe peak, $\nu_{\rm m} F_{\nu_{\rm m}}/\nu_{\pm, {\rm m}} F_{\nu_{\pm, {\rm m}}}$,
are well described by the ratio of the timescales given in Equation \eqref{eq:ratio_cooling_at_gamma_m}. This is true as long as $t_{\rm synch}$ is not much larger than $t_{\rm BeHe}$, so that the target photons for the BeHe process are those produced by proton synchrotron.

\begin{figure}
\centering
\begin{tabular}{cc}
\includegraphics[width=0.45\textwidth]{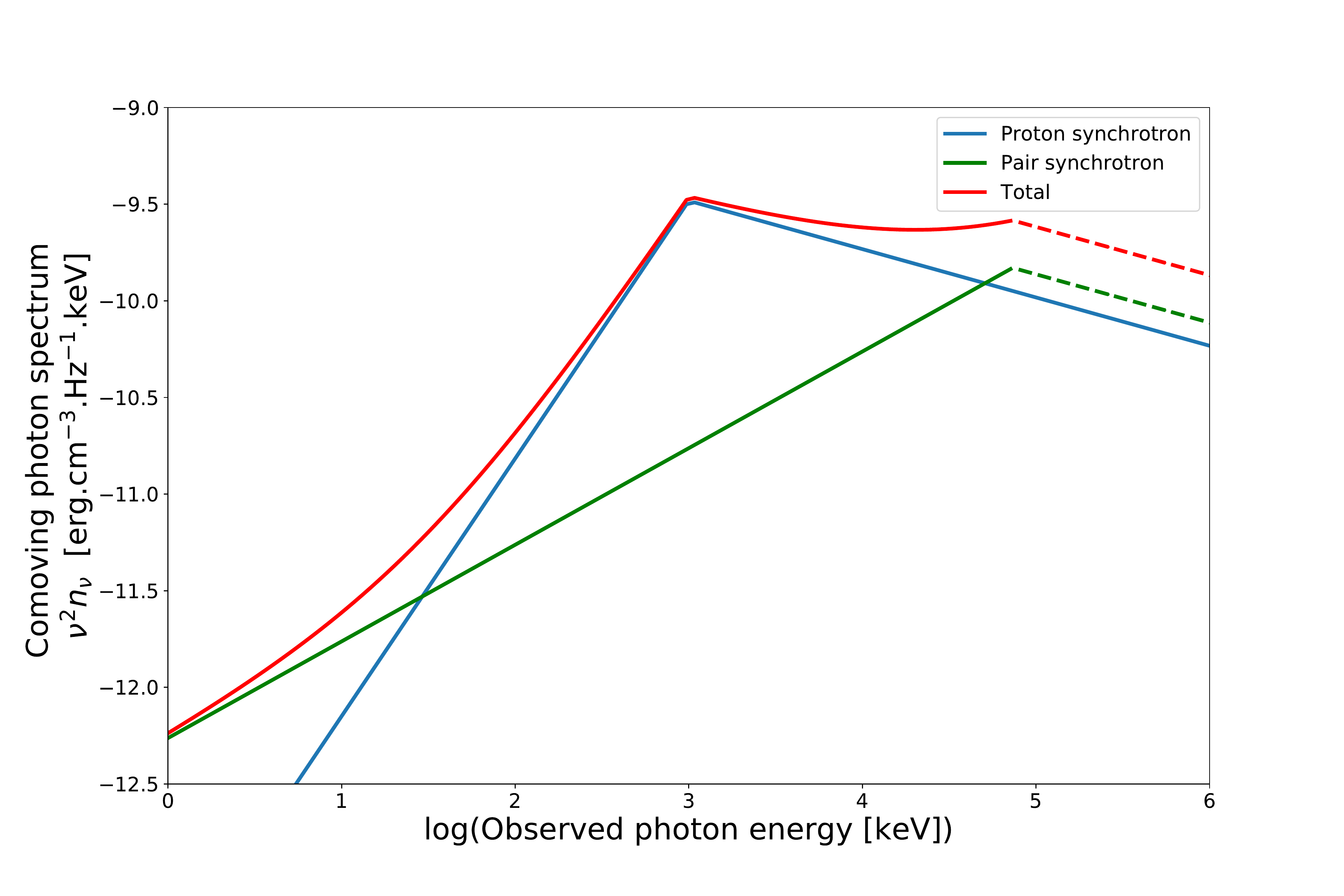} &
\includegraphics[width=0.45\textwidth]{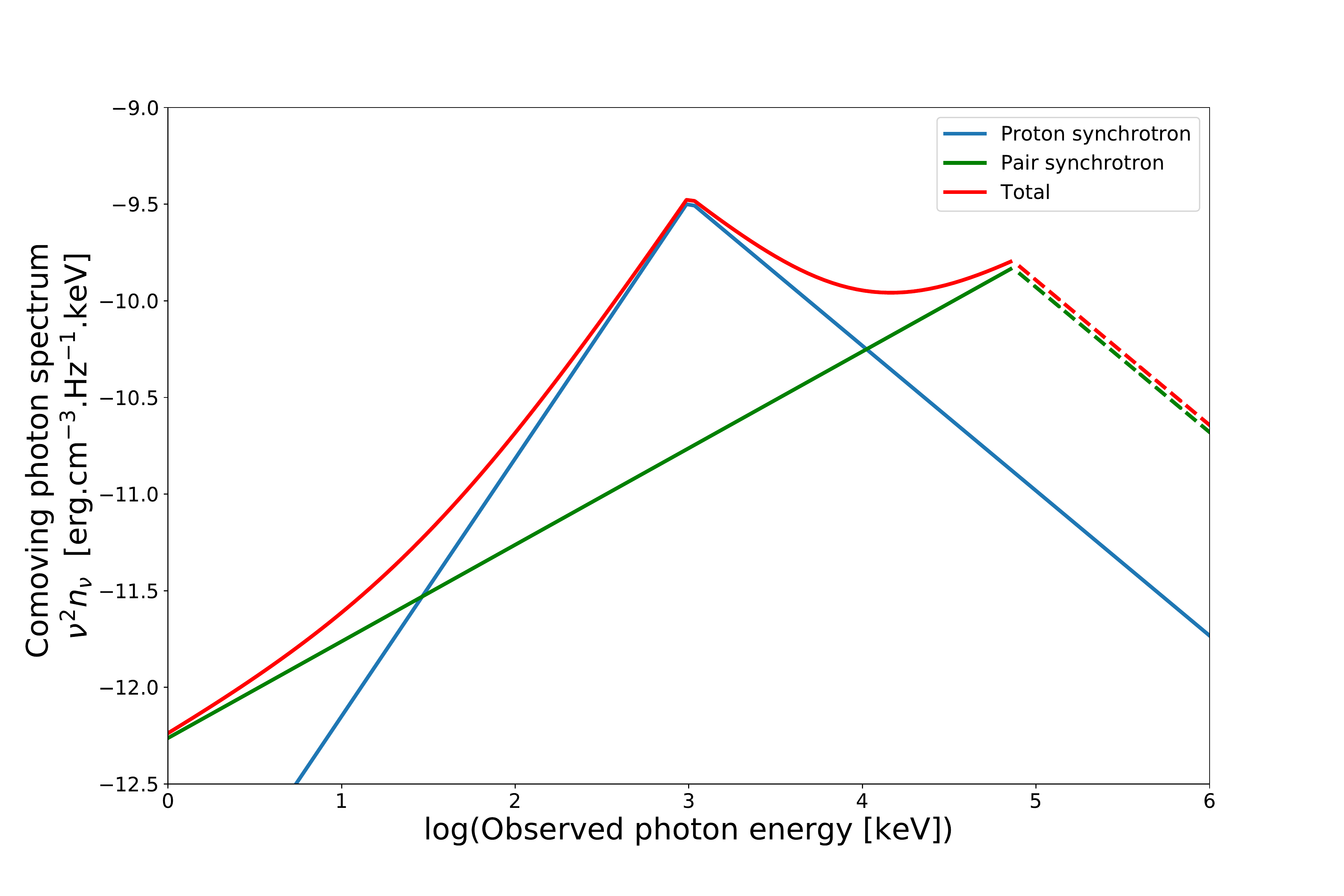}
\end{tabular}
\caption{Photon spectrum in marginally dominant synchrotron cooling regime with $t_{\rm synch} \sim t_{\rm BeHe}$. Parameters are $r_{14} = 1$, $\Gamma_2 = 1$, $\nu_{\rm MeV} = 1$, $L_{52} = 10$, $\xi = 1$ and $p = 2.5$ (left) or $p = 3.5$ (right). Blue - proton synchrotron component. Green - BeHe synchrotron component. Red - total. It is clear that the emission from the secondary BeHe pairs greatly affect the overall shape of the spectrum at low ($\lesssim 30$ keV) and high ($\gtrsim 10$ MeV) energies . The frequency of the second peak is independent on the power-law index. Above the photon peak at $\sim 100~$MeV, the photon spectrum is not specified by our analysis as it depends on the details of the pair injection spectrum, and as such on the exact shape of the proton distribution function above $\gamma_{\rm p,m}$. This is represented by a dashed line. }
\label{fig:spectrum_synchrotron_dominated}
\end{figure}

Figure \ref{fig:spectrum_synchrotron_dominated} shows an example spectrum when the two timescales are comparable.
A subdominant power-law extends from $\sim 100$ MeV all across the observation window. This extra component is
solely due to synchrotron radiation from the BeHe pairs and it is simultaneous to the main MeV emission.
It is potentially detectable at low energy (few tens of keV) and in the LLE data. In making this figure, we assumed
$r_{14} = 1$, $\Gamma_2 = 1$, $\nu_{\rm MeV} = 1$, $L_{52} = 10$, $\xi =1$ and $p = 2.5$ (left) or $p = 3.5$ (right). 
They both correspond to $t_{\rm BeHe}/ t_{\rm synch} \sim 1.9$ (the approximate expression in Equation
\eqref{eq:ratio_cooling_at_gamma_m} gives $t_{\rm BeHe}/ t_{\rm synch}\sim 0.85$). The figure was made with BeHe inelasticity
$\kappa_{\rm e} \sim 10^{-4}$ to determine $\nu_{{\rm m},\pm}$. This value of $\kappa_{\rm e}$ is appropriate
when $\gamma_{\rm p,m} h \nu_{\rm m}/(m_{\rm e} c^2) \sim 100 $, as obtained for this choice of parameters.
The total number of pairs was calculated using the integral formulation of \cite{CZS92} (see Appendix \ref{app:appendix_pair_production_rate_simplification}).

In this example, the overall synchrotron spectrum is strongly modified at low and high energies by an extra
power-law produced by the synchrotron emission of the BeHe pairs. This is a clear spectral signature of
the proton synchrotron emission model, which can help to differentiate proton synchrotron models from electron synchrotron models.
Several properties of this power-law are well determined and weakly sensitive to the parameter of
the model. First, its slope is set to be $\alpha = -1.5$
since it is produced by electrons and positrons in the fast cooling regime. Second, it extends from low (sub-keV)
energy to high energy with a peak at few tens or hundreds of MeV. Therefore, this component crosses the entire
GBM energy window. Third, the maximum energy of this component is linearly correlated with the peak frequency of the MeV
component, as shown by Equation \eqref{eq:ratio_peak_energy}. Finally, the intensity of this component does not
have a strong dependence on the uncertain value of the proton index $p$, but strongly depends on the luminosity, Lorentz factor and emission 
radius. The combination of all those characteristics provides a clear smoking-gun of proton synchrotron model. 

\section{Discussion and conclusion}
\label{sec:4}

Proton synchrotron models are an attractive solution to explain marginally fast cooling spectra from GRBs
\citep{GGO19}. 
We have performed a detailed investigation of the effect of BeHe cooling on the protons and of the subsequent
radiation of the BeHe pairs.  For high bulk Lorentz factor $\Gamma$, large radius $r$ and low luminosity
$L^{\rm obs}$, proton synchrotron emission dominates and no BeHe pair signature is expected. Conversely,
synchrotron emission from the BeHe pairs dominates when the luminosity is high, and $\Gamma$ and $r$ are relatively
low. This constitutes an additional test for the model: if no pair signature is observed in high
luminosity bursts, large radii and high Lorentz factors are necessary. The magnetic luminosity is given by
\begin{align}
L_{\rm B} = \pi c r^2 \Gamma^2 \frac{B^2}{8\pi} = 4.43 \times 10^{53} {\rm erg}~{\rm s}^{-1} r_{14}^{\frac{8}{3}} \xi^{\frac{4}{3}} t_{\rm v}^{-2} \nu_{\rm MeV}^{-\frac{2}{3}}
\label{eq:magnetic_luminosity}
\end{align}
where we have replaced the Lorentz factor $\Gamma$ by the variability time $t_{\rm v} = r/(c \Gamma^2)$ normalised
to 1s. The large dependence on the radius and variability time implies a very large magnetic luminosity if
no BeHe pair signature is observed.
For instance, for a variability timescale of the order a few seconds as observed in the burst sample of \cite{BBB19},
an observed luminosity of $L^{\rm obs} = 10^{54}~$erg s$^{-1}$ requires that $r\sim 10^{15}$
cm to suppress the BeHe pair creation (see Figure \ref{fig:cooling_comparison}). Using Equation
\eqref{eq:magnetic_luminosity}, this implies a magnetic luminosity of a few $10^{56}$ erg s$^{-1}$.
Such a high magnetic luminosity suggests that the jet is magnetically dominated, leading to an 
acceleration rate with $\Gamma \propto r^{1/3}$ \citep{DS02, BPL17}, and possibly magnetic reconnection as the
energy dissipation mechanism, see \textit{e.g.} \cite{LB01,GU19}.



In between the two regimes outlined above, we expect a parameter space where both signatures can be
observed simultaneously in the MeV band (see Figure \ref{fig:spectrum_synchrotron_dominated}).
This component might have already been observed in several GRBs. Indeed, it is
reminiscent of the population of bursts whose spectra seems to have two components: a main emission peak
together with a subdominant component well approximated by a power-law observed from a few keV to several
tens of MeV \citep{AAA10_Ferm_090510,GKD15}. This sub-dominant power-law can be interpreted as
the radiation from BeHe pairs, identified in our work as the spectral signature of proton synchrotron models.

Most notably, the spectra of GRB 190114C detected by MAGIC \citep{MAA20} are
composed of two components in the energy range 1keV - 1 GeV\footnote{Three when considering the TeV emission.},
with low energy slopes seemingly compatible with slow $\alpha = -2/3$ and fast $\alpha = -3/2$ cooling
synchrotron radiation in some time bins (\citet{CPB20}, however see \citet{AAA20_190114C}). In addition,
a spectral cut-off component at energies between $50$ to $100$ MeV was reported in \citet{CPB20}. We
speculate that this burst might possess the clear signature of proton synchrotron with BeHe cooling
discussed in this paper: 1) one main peak produced by proton synchrotron in the slow cooling regime,
2) a second component compatible with fast cooling synchrotron radiation from BeHe pairs and 3)
a cutoff between $50$--$100$ MeV corresponding to the injection limit of BeHe pairs. Furthermore,
analysis of panel c) and d) in Figure 2 of \citet{CPB20} shows that the ratio between the low
and the high peak frequency is about $100$, in rough agreement with Equation \eqref{eq:ratio_peak_energy}.


In our analysis, there are additional emission processes not dealt with that might modify the spectrum. We briefly discuss some of them here. First, we neglected the modification of the high energy peak by $\gamma\gamma$ absorption. Indeed, Equation \eqref{eq:ratio_peak_energy} shows that the synchrotron peak of the pairs is marginally below the energy threshold for pair creation. Therefore, only the spectrum at energies larger than the peak can be affected by this process. Pair recombination could produce an observable characteristic around observed frequency $\nu_{{\rm obs},\gamma \gamma} \sim 2\Gamma \, m_e c^2 /h = 2.4 \times 10^{22} ~{\rm Hz} ~ \Gamma_2$. Yet, the time for pair recombination approximated by  $t_{e^\pm \rightarrow \gamma\gamma} = 2 /(\sigma_{\rm T} n_\pm \langle v \rangle)$ is such that 
\begin{align}
    \frac{t_{e^\pm \rightarrow \gamma\gamma}}{t_{\rm dyn}} = 2.0 \times 10^2 ~ \Gamma_2^{\frac{10}{3}} r_{14}^{\frac{22}{9}} \kappa_{\rm e, -4} \nu_{\rm MeV}^{\frac{20}{9}} L_{52}^{-2},
\end{align}
where we have approximated the density of pairs as $n_\pm  = \dot n_\pm t_{\rm dyn}$ using the upper branch of Equation \eqref{eq:n_pm_dot} and use for the average electron velocity $\langle v \rangle = c$. Thus, pairs do not substantially recombine for our fiducial parameters. However, we note the large dependence on the parameters, therefore, such a signature could in some parameter space regions be present.

Furthermore, we did not treat the radiation from the initial population of electrons. This was discussed in
\citet{GGO19}, who argued that under the assumption that the same number of protons and electrons are
accelerated, the electrons would not be seen if they achieve the same injection spectrum as the protons since
their luminosity would be a factor $m_e/m_p$ lower. If instead of similar injection spectrum, both electrons
and protons carry the same energy, electrons would radiate their energy in the TeV band, which should trigger
a leptonic cascade. In addition, thermalization of the background electrons via synchrotron self-absorption
could also change the low energy spectrum.  We expect this process to change the spectrum mostly in the optical
band, therefore our conclusions would be largely unaffected as we have focused on the keV to GeV band.

We assumed that photo-pion interaction is inefficient in our model since the threshold for this process is not reached for the fiducial parameters, see Equation \eqref{eq:prod_gps_nup}. However, it was argued by \cite{FPM21} that cooling by photo-pion dominates cooling by the BeHe process. We show in Appendix \ref{app:photopioncooling} the additional conditions that the parameters should satisfy for BeHe cooling to dominate. We find that this holds true for a large set of parameters. Yet, if a substantial amount of energy is transferred from protons to charged pions by photo-hadronic interaction, the spectrum above the peak would be modified.

In a future work, we will compute a table model with \textit{SOPRANO}, a code designed to simulate
lepto-hadronic processes in optically thin environment \citep{GBS21}. There, we perform direct fits
of a proton synchrotron model with BeHe pair production to GRB 190114C, in order to understand if the
additional power-law and its cutoff are in agreement with the model presented here. The sample of
bursts in \cite{BBB19} will also be studied, in order to constrain the parameters and further test the model.

To conclude, proton synchrotron models have a clear smoking-gun signature: the synchrotron radiation
produced by the BeHe pairs. The emerging spectrum is composed of a main synchrotron peak produced by protons,
and a power-law with index $\alpha = -1.5$ extending from sub-keV energies to few tens or hundreds of MeV. The peak
frequency of this component is linearly linked to the frequency of the MeV peak frequency. Identification of this
extra power-law in spectra will help to constrain the emission mechanism of GRB jets and their parameters. 

\vspace{1cm}

\begin{acknowledgments}
DB and AP acknowledge support from the European Research Council via the ERC consolidating grant $\sharp$773062 (acronym O.M.J.).
F.S. acknowledges support from the G\"oran Gustafsson Foundation for Research in Natural Sciences and Medicine.
\end{acknowledgments}



\clearpage

\appendix

\section{Approximation to the Bethe-Heitler cooling rate}
\label{app:appendix_BeHe_cooling_rate_simplification}
The cooling rate for a proton of Lorentz factor $\gamma_{\rm p}$ is given in a simple form by \cite{CZS92}:
\begin{align}
-\frac{d\gamma_{\rm p}^{\rm  BeHe}}{dt} = \alpha r_0^2 c \frac{m_{\rm e}}{m_{\rm p}} \int_2^\infty d\kappa n_x\left( \frac{\kappa}{2\gamma_{\rm p}} \right ) \frac{\phi(\kappa)}{\kappa^2}, \label{app_eq:dgp_dt}
\end{align}
where $\alpha$ is the fine structure constant, $r_0$ the classical electron radius, $\kappa = 2\gamma_{\rm p} x$ is the maximum energy for a photon with energy $x = h\nu/m_ec^2$ in the proton rest frame, and $\phi(\kappa)/\kappa^2$ is the energy loss rate of a single-energy proton with Lorentz factor $\gamma_p$ in an isotropic photon background. It is approximately given by Equations (3.14) and (3.18) of \cite{CZS92}. The photon distribution $n_x$ is the photon number per energy in units electron rest mass. In Equation \eqref{app_eq:dgp_dt}, the photon distribution function is used at the value $\kappa/(2\gamma_{\rm p})$. It is given by (compare with Equation \eqref{eq:synch_spectrum_fast_cooling})
\begin{align}
n_x (x) = \frac{u_{\nu_{\rm m}}}{hx} \left\{
\begin{aligned}
		& \left(\frac{x}{x_{\rm c}}\right)^{1/3}  \xi^{1/2} \quad &&x < x_{\rm c}\\
		&\left(\frac{x}{x_{\rm m}}\right)^{-1/2}  \quad &&x_{\rm c} < x < x_{\rm m}\\
		&\left(\frac{x}{x_{\rm m}}\right)^{-p/2}  \quad &&x > x_{\rm m}
\end{aligned}
\right.\label{eq:app_photon_spectrum_mec2}
\end{align}
where $x_{\rm c} = h\nu_c/(m_{\rm e} c^2)$ and $x_{\rm m} = h\nu_{\rm m}/ m_{\rm e}c^2$ are the cooling and injection frequencies in units of the electron rest mass energy, $\xi = \nu_c / \nu_m$ and $u_{\nu_m}$ is the peak spectral energy density as given by Equation \eqref{eq:spectral_energy_density_synchrotron_proton}.
In order to obtain analytical results that still factorize the photon spectrum, we approximate $\phi(\kappa)/\kappa^2$ by a $\delta$-function that is normalized to the integral between $2 < \kappa < 500 $ (those numbers are arbitrarily, but we numerically checked that they encompass most of the cooling contribution):
\begin{align}
\frac{\phi(\kappa)}{\kappa^2} = A_{\pm} \delta(\kappa - \kappa_0) 
\end{align}
where we find $A_\pm = 420$ and $\kappa_0 = 40$, roughly corresponding to the position of the maximum of $\phi(\kappa)/\kappa^2$.
Using this approximation in Equation \eqref{app_eq:dgp_dt} gives
\begin{align}
-\frac{d\gamma_{\rm p}^{\rm  BeHe}}{dt} \sim \alpha r_0^2 c \frac{m_{\rm e}}{m_{\rm p}} A_\pm n_x \left ( \frac{\kappa_0}{2\gamma_{\rm p}}\right )
\end{align}
We now use the expression for the photon spectrum given in Equation \eqref{eq:synch_spectrum_fast_cooling} to obtain in 
marginally fast ($\xi > 1$) cooling 
\begin{align}
-\frac{d\gamma_{\rm p}^{\rm BeHe}}{dt} =& \alpha r_0^2 c \frac{m_{\rm e}}{m_{\rm p}} \frac{A_\pm}{h } u_{\nu_{\rm m}}  \left \{ \begin{aligned}
& \left ( \frac{2 \gamma_{\rm p}}{\kappa_0} \right)^{\frac{2}{3}} \left ( \frac{1}{x_{\rm c}} \right )^{\frac{1}{3}} \xi^{\frac{1}{2}} & & \frac{\kappa_0}{2\gamma_{\rm p}} < x_{\rm c}\\
& \left ( \frac{2\gamma_{\rm p}}{\kappa_0} \right )^{\frac{3}{2}} \left ( \frac{1}{x_{\rm m}} \right )^{-\frac{1}{2}} & & x_{\rm c} < \frac{\kappa_0}{2\gamma_{\rm p}} < x_{\rm m} \\
& \left ( \frac{2 \gamma_{\rm p}}{\kappa_0} \right )^{1+\frac{p}{2}} \left ( \frac{1}{x_{\rm m}} \right )^{-\frac{p}{2}} & & x_{\rm m} <  \frac{\kappa_0}{2\gamma_{\rm p}} 
\end{aligned} \right. \label{app:eq_dgammadt_BeHe}
\end{align}

The cooling time by the BeHe process is finally obtained by 
\begin{align}
t_{\rm BeHe}^{-1} = \left | \frac{1}{\gamma_{\rm p}}\frac{d\gamma_{\rm p}^{\rm BeHe}}{dt} \right | \label{app:eq_BeHe_cooling}
\end{align}

\section{Bethe-Heitler pair production rate in marginally fast cooling}
\label{app:appendix_pair_production_rate_simplification}

In this appendix, we seek an approximate expression for the pair production under the assumption of marginally fast cooling $\xi \equiv 1$. The pair production rate is given by \cite{CZS92}
\begin{align}
\frac{\partial N_\pm}{\partial t} = c \frac{3}{8\pi} \frac{\sigma_{\rm T} \alpha}{\gamma_{\rm p}} \int_2^\infty n_{ x} \left ( \frac{\kappa}{2\gamma_{\rm p}}\right ) \frac{\psi(\kappa)}{\kappa^2}   d\kappa, \label{eq:pair_prod_rate_CZS92}
\end{align}
where $\psi(\kappa)$ is given by Equations 2.3 of \citet{CZS92}. Multiplying this equation by an electron energy $\gamma_\pm m_e c^2$ and integrating over electron energies corresponds to the cooling rate of a proton whose expression is given by Equation \eqref{app_eq:dgp_dt}.
Replacing with the expression of the synchrotron spectrum in the marginally fast cooling scenario with $\gamma_{\rm m} = \gamma_{\rm c}$ gives:
\begin{align}
\frac{8\pi \gamma_{\rm p}}{3 c \sigma_{\rm T} \alpha} \frac{\partial N_\pm}{\partial t} = \int_2^{\kappa_*} n_x\left ( \frac{\kappa}{2\gamma_{\rm p}}\right ) \frac{\psi(\kappa)}{\kappa^2}   d\kappa +\int_{\kappa_*}^\infty n_x\left ( \frac{\kappa}{2\gamma_{\rm p}}\right ) \frac{\psi(\kappa)}{\kappa^2}   d\kappa
\end{align}
where $\kappa_* \geq 2$ is such that
\begin{align}
\frac{\kappa_*}{2\gamma_{\rm p}} = \frac{h\nu_{\rm m}}{m_{\rm e} c^2}
\end{align}
is the energy of a photon at the spectral peak in the rest frame of the proton. Using the expression for the photon spectrum given by Equation \eqref{eq:synch_spectrum_fast_cooling}, it becomes:
\begin{align}
\frac{8 \pi \gamma_{\rm p}}{ 3 c \sigma_{\rm T} \alpha} \frac{\partial N_\pm}{\partial t} & =  \frac{2\gamma_{\rm p}}{h} \left ( \frac{1 }{2\gamma_{\rm p} x_{\rm m}}\right )^{\frac{1}{3}} u_{\nu_{\rm m}} \int_2^{\kappa_*} \kappa^{-\frac{2}{3}}  \frac{\psi(\kappa)}{\kappa^2}   d\kappa + \frac{2\gamma_{\rm p}}{h} \left( \frac{1}{2\gamma_{\rm p} x_{\rm m}} \right)^{-\frac{p}{2}} u_{\nu_{\rm m}} \int_{\kappa_*}^\infty  \kappa^{-\frac{p_{\rm e}}{2}-1} \frac{\psi(\kappa)}{\kappa^2}   d\kappa  \label{eq:simplified_pair_prod_rate}
\end{align}

\section{Electron spectrum for $\delta-$function injection with synchrotron cooling}
\label{sec:appendix_pair_spectrum}

We assume that electron-positron pairs are produced with a single Lorentz factor $\gamma_{\pm,{\rm m}}$ and we proceed by solving the kinetic equation describing the evolution of the pair distribution function 
\begin{align}
\frac{\partial n_\pm}{\partial t} + \frac{1}{m_{\rm e} c^2} \frac{\partial }{\partial \gamma_{\rm e}} \left ( P_{\rm e}(\gamma) n_\pm \right ) = \dot n_\pm \delta(\gamma - \gamma_{\pm,{\rm m}})
\end{align}
where the pair distribution function $n_\pm(\gamma_{\rm e},t)$ is a function of both time and electron energy, and $\dot n_\pm$ is the pair production rate by the BeHe process, given by Equation \eqref{eq:pair_production_rate}. Let $\epsilon $ be a small positive variable and let us integrate the above equation between  $\gamma_{\pm,{\rm m}} - \epsilon$ and $\gamma_{\pm,{\rm m}} +\epsilon$. Keeping only the zeroth order term in $\epsilon$, we obtain
\begin{align}
\frac{1}{m_{\rm e} c^2} \left [ P_{\rm e} n_\pm \right ]_{\gamma_{\pm,{\rm m}} -\epsilon}^{\gamma_{\pm,{\rm m}} +\epsilon} = \dot n_\pm \label{eq:app_jump_gpm}
\end{align}
describing a jump in the solution at $\gamma_{\pm,{\rm m}}$. Since particles are cooling, $n_\pm$ is null for $\gamma > \gamma_{\pm,{\rm m}}$. We now solve the equation for $\gamma < \gamma_{\pm,{\rm m}}$. Further assuming that electron are fast cool, far from $\gamma_{\pm, \rm c} \ll \gamma_{\pm,{\rm m}}$, the equation can be further simplified to
\begin{align}
\frac{ \partial}{\partial \gamma} \left ( P_{\rm e} n_\pm \right ) = 0
\end{align}
Therefore, the solution is of the form
\begin{align}
n_\pm = Q \times \left(\frac{\gamma_{\pm,{\rm m}}}{\gamma} \right )^2 \label{app_eq:pair_spectrum_delta}
\end{align}
where we introduced $\gamma_{\pm,{\rm m}}$ for convenience. We use Equation \eqref{eq:app_jump_gpm} to obtain 
\begin{align}
Q = \frac{m_{\rm e} c^2}{P_{\rm e}(\gamma_{\pm,{\rm m}})} \dot n_\pm
\end{align}

\section{Photo-pion cooling of protons on proton synchrotron photons}
\label{app:photopioncooling}

The cooling of protons by photopion interaction is given by \citep{MK94,BRS90,WB97}
\begin{align}
t_{\rm p \pi}^{-1} = \frac{c}{2\gamma_{\rm p}^2} \int_{\bar \epsilon_{\rm th}}^\infty d\bar \epsilon \sigma_{p\pi} (\bar \epsilon) K_p (\bar \epsilon) \bar \epsilon \int_{\frac{\bar \epsilon}{2 \gamma_{\rm p}}}^{\infty} dx \frac{n_x(x)}{x^2} 
\end{align}
Following \cite{PM15b},  we set
\begin{align}
\sigma_{\rm p\pi} &= \sigma_0 H(\bar \epsilon - \bar \epsilon_{\rm th}) \\
\sigma_0 &= 1.5 \times 10^{-4} \sigma_{\rm T} \\
K_{\rm p} &= 0.2 \\
\bar \epsilon_{\rm th} &= 145 {\rm ~MeV}
\end{align}
where $H$ is the Heaviside function. The cooling time becomes 
\begin{align}
t_{\rm p \pi}^{-1} = \frac{c}{2\gamma_{\rm p}^2} K_p \sigma_0 \int_{\bar \epsilon_{\rm th}}^\infty d\bar \epsilon \bar \epsilon \int_{\frac{\bar \epsilon}{2 \gamma_{\rm p}}}^{\infty} dx \frac{n(x)}{x^2} ,
\end{align} 
where $\bar \epsilon$ and $\bar \epsilon_{\rm th}$ are expressed in units of the electron rest mass energy.

Using the photon distribution function given by Equation \eqref{eq:app_photon_spectrum_mec2}, assuming $x_m = x_c$ and $2\gamma_{\rm p} x_m  > \bar \epsilon_{\rm th}$, we obtain
\begin{align}
\frac{t_{\rm p \pi}^{-1}}{ \frac{c}{2\gamma_{\rm p}^2} K_p \sigma_0} = & \int_{\bar \epsilon_{\rm th}}^{2\gamma_p x_m} d\bar \epsilon \bar \epsilon \left [ \int_{\frac{\bar \epsilon}{2 \gamma_{\rm p}}}^{x_m} dx \frac{1}{x^2} \frac{1}{h} \frac{1}{x} u_{\nu_m}  \left(\frac{x}{x_m} \right )^{\frac{1}{3}} + \int_{x_m}^{\infty} dx \frac{1}{x^2} \frac{1}{h} \frac{1}{x} u_{\nu_m} \left ( \frac{x}{x_m}\right )^{-\frac{p}{2}}\right ]   \\ 
+ & \int_{2 \gamma_p x_m}^\infty d\bar \epsilon \bar \epsilon \int_{\frac{\bar \epsilon}{2 \gamma_{\rm p}}}^{\infty} dx \frac{1}{x^2} \frac{1}{h} \frac{1}{x} u_{\nu_m} \left ( \frac{x}{x_m}\right )^{-\frac{p}{2}} \nonumber 
\end{align}
After some algebra, it simplifies to
\begin{align}
\frac{t_{\rm p \pi}^{-1}}{ \frac{c}{2\gamma_{\rm p}^2} K_p \sigma_0 \frac{u_{\nu_m}}{h} } = \left \{ \begin{aligned}
& \frac{(2\gamma_p)^{2}}{(2+\frac{p}{2})\frac{p}{2}}   & ~~~~~~~ & 2 \gamma_p x_m \rightarrow \bar \epsilon_{\rm th}^+ \\
& (2 \gamma_p)^2 \left [ \frac{9}{5}   +  \left ( \frac{1}{(2 + \frac{p}{2})} - \frac{1}{(5/3) } \right ) \frac{1}{2}    + \frac{1}{(2+\frac{p}{2})\frac{p}{2}} \right ] & & 2 \gamma_p x_m \gg \bar \epsilon_{\rm th}
\end{aligned} \right .
\end{align}
where the dependence on the energy $x_{m}$ simplifies with the normalisation of the photon spectrum.
If $2 \gamma_p x_m < \bar \epsilon_{\rm th} $, the cooling time is simplified as
\begin{align}
\frac{t_{\rm p \pi}^{-1}}{ \frac{c}{2\gamma_{\rm p}^2} K_p \sigma_0 \frac{u_{\nu_m}}{h}} = &  \left (\frac{x_m}{\bar \epsilon_{\rm th}} \right )^{\frac{p}{2}} \frac{(2\gamma_p)^{2 + \frac{p}{2}}}{ \left (2 +\frac{p}{2} \right ) \frac{p}{2} } 
\end{align}

To compare BeHe and photopion cooling, only three cases needs to be considered : they depend on the location of $2 \gamma_{p} x_m$ with respect to $\kappa_0$ and $\bar \epsilon_{\rm th}$. The condition $2 x_m \gamma_{\rm p,m} > \bar \epsilon_{\rm th} > \kappa_0$ is rarely satisfied. If it is, photo-pion cooling dominates by large over BeHe cooling.
For $\kappa_0 < 2\gamma_{\rm p} x_{\rm m}  < \bar \epsilon_{\rm th}$, the ratio of cooling time is simplified to
\begin{align}
\frac{t_{\rm BeHe}^{-1}}{t_{\rm p\pi}^{-1}} = \left (2 +\frac{p}{2} \right ) \frac{p}{2} \frac{A_\pm}{K_p} \frac{\bar \epsilon_{\rm th} ^{\frac{p}{2}}}{\kappa_0^{\frac{2}{3}}  } \frac{  \alpha r_0^2 \frac{m_{\rm e}}{m_{\rm p}}   }{  \sigma_0} \frac{1}{  x_m^{\frac{p}{2} + \frac{1}{3}} (2\gamma_p)^{\frac{p}{2} + \frac{1}{3}}  } = 0.39 \Gamma_2^{p + \frac{2}{3}} \nu_{1 MeV} ^{-\frac{5p}{6} - \frac{5}{9}} R_{14}^{-\frac{p}{6} - \frac{1}{9}}
\end{align}
where we set $\nu_c = \nu_m$ (i.e. $\xi = 1$). For the computation of the numerical factor, we set $p = 2.5$. In that case, photo-pion cooling dominates over BeHe, but we point the strong dependence on the Lorentz factor $\Gamma$. For $\Gamma_2 = 3$, the cooling is dominated by BeHe.
In the case $\bar \epsilon_{\rm th} > \kappa_0 > 2\gamma_{\rm p} x_{\rm m}$, it comes
\begin{align}
\frac{t_{\rm BeHe}^{-1}}{t_{\rm p\pi}^{-1}}  & = \left (2 +\frac{p}{2} \right ) \frac{p}{2} \frac{A_\pm}{K_p} \frac{\bar \epsilon_{\rm th}^{\frac{p}{2}}}{\kappa_0^{1+\frac{p}{2}}} \frac{ \alpha r_0^2 \frac{m_{\rm e}}{m_{\rm p}}  }{  \sigma_0 }  \sim 7.7,
\end{align}
which is independent on $\gamma_p$ and on $x_m$, and therefore remains constant for any parameters of the problem. In this numerical estimate, we set $p = 2.5$.

\bibliography{biblio.bib}

\begin{thebibliography}{}
\expandafter\ifx\csname natexlab\endcsname\relax\def\natexlab#1{#1}\fi
\providecommand{\url}[1]{\href{#1}{#1}}
\providecommand{\dodoi}[1]{doi:~\href{http://doi.org/#1}{\nolinkurl{#1}}}
\providecommand{\doeprint}[1]{\href{http://ascl.net/#1}{\nolinkurl{http://ascl.net/#1}}}
\providecommand{\doarXiv}[1]{\href{https://arxiv.org/abs/#1}{\nolinkurl{https://arxiv.org/abs/#1}}}

\bibitem[{Aartsen {et~al.}(2017)Aartsen, Ackermann, Adams, Aguilar, Ahlers,
  Ahrens, Samarai, Altmann, Andeen, Anderson, {et~al.}}]{AAA17}
Aartsen, M., Ackermann, M., Adams, J., {et~al.} 2017, \apjl, 843, 112A

\bibitem[{{Ackermann} {et~al.}(2010){Ackermann}, {Asano}, {Atwood}, {Axelsson},
  {Baldini}, {Ballet}, {Barbiellini}, {Baring}, {Bastieri}, {Bechtol},
  {Bellazzini}, {Berenji}, {Bhat}, {Bissaldi}, {Blandford}, {Bloom},
  {Bonamente}, {Borgland}, {Bouvier}, {Bregeon}, {Brez}, {Briggs}, {Brigida},
  {Bruel}, {Buson}, {Caliandro}, {Cameron}, {Caraveo}, {Carrigan},
  {Casandjian}, {Cecchi}, {{\c C}elik}, {Charles}, {Chiang}, {Ciprini},
  {Claus}, {Cohen-Tanugi}, {Connaughton}, {Conrad}, {Dermer}, {de Palma},
  {Dingus}, {Silva}, {Drell}, {Dubois}, {Dumora}, {Farnier}, {Favuzzi},
  {Fegan}, {Finke}, {Focke}, {Frailis}, {Fukazawa}, {Fusco}, {Gargano},
  {Gasparrini}, {Gehrels}, {Germani}, {Giglietto}, {Giordano}, {Glanzman},
  {Godfrey}, {Granot}, {Grenier}, {Grondin}, {Grove}, {Guiriec}, {Hadasch},
  {Harding}, {Hays}, {Horan}, {Hughes}, {J{\'o}hannesson}, {Johnson}, {Kamae},
  {Katagiri}, {Kataoka}, {Kawai}, {Kippen}, {Kn{\"o}dlseder}, {Kocevski},
  {Kouveliotou}, {Kuss}, {Lande}, {Latronico}, {Lemoine-Goumard}, {Llena
  Garde}, {Longo}, {Loparco}, {Lott}, {Lovellette}, {Lubrano}, {Makeev},
  {Mazziotta}, {McEnery}, {McGlynn}, {Meegan}, {M{\'e}sz{\'a}ros}, {Michelson},
  {Mitthumsiri}, {Mizuno}, {Moiseev}, {Monte}, {Monzani}, {Moretti},
  {Morselli}, {Moskalenko}, {Murgia}, {Nakajima}, {Nakamori}, {Nolan},
  {Norris}, {Nuss}, {Ohno}, {Ohsugi}, {Omodei}, {Orlando}, {Ormes}, {Ozaki},
  {Paciesas}, {Paneque}, {Panetta}, {Parent}, {Pelassa}, {Pepe},
  {Pesce-Rollins}, {Piron}, {Preece}, {Rain{\`o}}, {Rando}, {Razzano},
  {Razzaque}, {Reimer}, {Ritz}, {Rodriguez}, {Roth}, {Ryde}, {Sadrozinski},
  {Sander}, {Scargle}, {Schalk}, {Sgr{\`o}}, {Siskind}, {Smith}, {Spandre},
  {Spinelli}, {Stamatikos}, {Stecker}, {Strickman}, {Suson}, {Tajima},
  {Takahashi}, {Takahashi}, {Tanaka}, {Thayer}, {Thayer}, {Thompson},
  {Tibaldo}, {Toma}, {Torres}, {Tosti}, {Tramacere}, {Uchiyama}, {Uehara},
  {Usher}, {van der Horst}, {Vasileiou}, {Vilchez}, {Vitale}, {von Kienlin},
  {Waite}, {Wang}, {Wilson-Hodge}, {Winer}, {Wu}, {Yamazaki}, {Yang}, {Ylinen},
  \& {Ziegler}}]{AAA10_Ferm_090510}
{Ackermann}, M., {Asano}, K., {Atwood}, W.~B., {et~al.} 2010, \apj, 716, 1178,
  \dodoi{10.1088/0004-637X/716/2/1178}

\bibitem[{{Acuner} {et~al.}(2020){Acuner}, {Ryde}, {Pe'er}, {Mortlock}, \&
  {Ahlgren}}]{ARP20}
{Acuner}, Z., {Ryde}, F., {Pe'er}, A., {Mortlock}, D., \& {Ahlgren}, B. 2020,
  \apj, 893, 128, \dodoi{10.3847/1538-4357/ab80c7}

\bibitem[{{Ahlgren} {et~al.}(2015){Ahlgren}, {Larsson}, {Nymark}, {Ryde}, \&
  {Pe'er}}]{ALN15}
{Ahlgren}, B., {Larsson}, J., {Nymark}, T., {Ryde}, F., \& {Pe'er}, A. 2015,
  \mnras, 454, L31, \dodoi{10.1093/mnrasl/slv114}

\bibitem[{{Ajello} {et~al.}(2020){Ajello}, {Arimoto}, {Axelsson}, {Baldini},
  {Barbiellini}, {Bastieri}, {Bellazzini}, {Berretta}, {Bissaldi}, {Blandford},
  {Bonino}, {Bottacini}, {Bregeon}, {Bruel}, {Buehler}, {Burns}, {Buson},
  {Cameron}, {Caputo}, {Caraveo}, {Cavazzuti}, {Chen}, {Chiaro}, {Ciprini},
  {Cohen-Tanugi}, {Costantin}, {Cutini}, {D'Ammando}, {DeKlotz}, {de la Torre
  Luque}, {de Palma}, {Desai}, {Di Lalla}, {Di Venere}, {Fana Dirirsa},
  {Fegan}, {Franckowiak}, {Fukazawa}, {Funk}, {Fusco}, {Gargano}, {Gasparrini},
  {Giglietto}, {Gill}, {Giordano}, {Giroletti}, {Granot}, {Green}, {Grenier},
  {Grondin}, {Guiriec}, {Hays}, {Horan}, {J{\'o}hannesson}, {Kocevski},
  {Kovac'evic'}, {Kuss}, {Larsson}, {Latronico}, {Lemoine-Goumard}, {Li},
  {Liodakis}, {Longo}, {Loparco}, {Lovellette}, {Lubrano}, {Maldera},
  {Malyshev}, {Manfreda}, {Mart{\'\i}-Devesa}, {Mazziotta}, {McEnery}, {Mereu},
  {Meyer}, {Michelson}, {Mitthumsiri}, {Mizuno}, {Monzani}, {Moretti},
  {Morselli}, {Moskalenko}, {Negro}, {Nuss}, {Omodei}, {Orienti}, {Orlando},
  {Palatiello}, {Paliya}, {Paneque}, {Pei}, {Persic}, {Pesce-Rollins},
  {Petrosian}, {Piron}, {Poon}, {Porter}, {Principe}, {Racusin}, {Rain{\`o}},
  {Rando}, {Rani}, {Razzano}, {Razzaque}, {Reimer}, {Reimer}, {Ryde}, {Saz
  Parkinson}, {Serini}, {Sgr{\`o}}, {Siskind}, {Spandre}, {Spinelli}, {Tajima},
  {Takagi}, {Takahashi}, {Tak}, {Thayer}, {Thompson}, {Torres}, {Troja},
  {Valverde}, {Van Klaveren}, {Wood}, {Yassine}, {Zaharijas}, {Mailyan},
  {Bhat}, {Briggs}, {Cleveland}, {Giles}, {Goldstein}, {Hui}, {Malacaria},
  {Preece}, {Roberts}, {Veres}, {Wilson-Hodge}, {Kienlin}, {Cenko}, {O'Brien},
  {Beardmore}, {Lien}, {Osborne}, {Tohuvavohu}, {D'Elia}, {D'A{\`\i}}, {Perri},
  {Gropp}, {Klingler}, {Capalbi}, {Tagliaferri}, {Stamatikos}, \& {De
  Pasquale}}]{AAA20_190114C}
{Ajello}, M., {Arimoto}, M., {Axelsson}, M., {et~al.} 2020, \apj, 890, 9,
  \dodoi{10.3847/1538-4357/ab5b05}

\bibitem[{{Albert} {et~al.}(2017){Albert}, {Andr{\'e}}, {Anghinolfi}, {Anton},
  {Ardid}, {Aubert}, {Avgitas}, {Baret}, {Barrios-Mart\'{\i}}, {Basa},
  {Bertin}, {Biagi}, {Bormuth}, {Bourret}, {Bouwhuis}, {Bruijn}, {Brunner},
  {Busto}, {Capone}, {Caramete}, {Carr}, {Celli}, {Chiarusi}, {Circella},
  {Coelho}, {Coleiro}, {Coniglione}, {Costantini}, {Coyle}, {Creusot},
  {Deschamps}, {De Bonis}, {Distefano}, {di Palma}, {Donzaud}, {Dornic},
  {Drouhin}, {Eberl}, {El Bojaddaini}, {Els{\"a}sser}, {Enzenh{\"o}fer},
  {Felis}, {Fusco}, {Galat{\`a}}, {Gay}, {Gei{\ss}els{\"o}der}, {Geyer},
  {Giordano}, {Gleixner}, {Glotin}, {Gregoire}, {Gracia-Ruiz}, {Graf},
  {Hallmann}, {van Haren}, {Heijboer}, {Hello}, {Hern{\'a}ndez-Rey},
  {H{\"o}{\ss}l}, {Hofest{\"a}dt}, {Hugon}, {Illuminati}, {James}, {de Jong},
  {Jongen}, {Kadler}, {Kalekin}, {Katz}, {Kie{\ss}ling}, {Kouchner}, {Kreter},
  {Kreykenbohm}, {Kulikovskiy}, {Lachaud}, {Lahmann}, {Lef{\`e}vre}, {Leonora},
  {Lotze}, {Loucatos}, {Marcelin}, {Margiotta}, {Marinelli},
  {Mart\'{\i}nez-Mora}, {Mathieu}, {Mele}, {Melis}, {Michael}, {Migliozzi},
  {Moussa}, {Mueller}, {Nezri}, {P{\u a}v{\u a}la}, {Pellegrino}, {Perrina},
  {Piattelli}, {Popa}, {Pradier}, {Quinn}, {Racca}, {Riccobene}, {Roensch},
  {S{\'a}nchez-Losa}, {Salda{\~n}a}, {Salvadori}, {Samtleben}, {Sanguineti},
  {Sapienza}, {Schnabel}, {Sch{\"u}ssler}, {Seitz}, {Sieger}, {Spurio},
  {Stolarczyk}, {Taiuti}, {Tayalati}, {Trovato}, {Tselengidou}, {Turpin},
  {T{\"o}nnis}, {Vallage}, {Vall{\'e}e}, {Van Elewyck}, {Vivolo}, {Vizzocca},
  {Wagner}, {Wilms}, {Zornoza}, \& {Z{\'u}{\~n}iga}}]{AAA17_Antares}
{Albert}, A., {Andr{\'e}}, M., {Anghinolfi}, M., {et~al.} 2017, \mnras, 469,
  906, \dodoi{10.1093/mnras/stx902}

\bibitem[{{Asano} {et~al.}(2009){Asano}, {Inoue}, \&
  {M{\'e}sz{\'a}ros}}]{AIM09}
{Asano}, K., {Inoue}, S., \& {M{\'e}sz{\'a}ros}, P. 2009, \apj, 699, 953,
  \dodoi{10.1088/0004-637X/699/2/953}

\bibitem[{{Axelsson} \& {Borgonovo}(2015)}]{AB15}
{Axelsson}, M., \& {Borgonovo}, L. 2015, \mnras, 447, 3150,
  \dodoi{10.1093/mnras/stu2675}

\bibitem[{{Band} {et~al.}(1993){Band}, {Matteson}, {Ford}, {Schaefer},
  {Palmer}, {Teegarden}, {Cline}, {Briggs}, {Paciesas}, {Pendleton}, {Fishman},
  {Kouveliotou}, {Meegan}, {Wilson}, \& {Lestrade}}]{BMF93}
{Band}, D., {Matteson}, J., {Ford}, L., {et~al.} 1993, \apj, 413, 281,
  \dodoi{10.1086/172995}

\bibitem[{{Bednarz} \& {Ostrowski}(1998)}]{BO98}
{Bednarz}, J., \& {Ostrowski}, M. 1998, \prl, 80, 3911,
  \dodoi{10.1103/PhysRevLett.80.3911}

\bibitem[{{Begelman} {et~al.}(1990){Begelman}, {Rudak}, \& {Sikora}}]{BRS90}
{Begelman}, M.~C., {Rudak}, B., \& {Sikora}, M. 1990, \apj, 362, 38,
  \dodoi{10.1086/169241}

\bibitem[{{B{\'e}gu{\'e}} {et~al.}(2017){B{\'e}gu{\'e}}, {Pe'er}, \&
  {Lyubarsky}}]{BPL17}
{B{\'e}gu{\'e}}, D., {Pe'er}, A., \& {Lyubarsky}, Y. 2017, \mnras, 467, 2594,
  \dodoi{10.1093/mnras/stx237}

\bibitem[{{B{\'e}gu{\'e}} {et~al.}(2013){B{\'e}gu{\'e}}, {Siutsou}, \&
  {Vereshchagin}}]{BSV13}
{B{\'e}gu{\'e}}, D., {Siutsou}, I.~A., \& {Vereshchagin}, G.~V. 2013, \apj,
  767, 139, \dodoi{10.1088/0004-637X/767/2/139}

\bibitem[{{B{\'e}gu{\'e}} \& {Vereshchagin}(2014)}]{BV14}
{B{\'e}gu{\'e}}, D., \& {Vereshchagin}, G.~V. 2014, \mnras, 439, 924,
  \dodoi{10.1093/mnras/stu011}

\bibitem[{{Beloborodov}(2010)}]{Bel10}
{Beloborodov}, A.~M. 2010, \mnras, 407, 1033,
  \dodoi{10.1111/j.1365-2966.2010.16770.x}

\bibitem[{{Beniamini} {et~al.}(2018){Beniamini}, {Barniol Duran}, \&
  {Giannios}}]{BBG18}
{Beniamini}, P., {Barniol Duran}, R., \& {Giannios}, D. 2018, \mnras, 476,
  1785, \dodoi{10.1093/mnras/sty340}

\bibitem[{{B{\"o}ttcher} \& {Dermer}(1998)}]{BD98}
{B{\"o}ttcher}, M., \& {Dermer}, C.~D. 1998, \apjl, 499, L131,
  \dodoi{10.1086/311366}

\bibitem[{{Bo{\v s}njak} {et~al.}(2009){Bo{\v s}njak}, {Daigne}, \&
  {Dubus}}]{BDB09}
{Bo{\v s}njak}, {\v Z}., {Daigne}, F., \& {Dubus}, G. 2009, \aap, 498, 677,
  \dodoi{10.1051/0004-6361/200811375}

\bibitem[{Burgess(2019)}]{Bur17}
Burgess, J.~M. 2019, \aap, 629, A69, \dodoi{10.1051/0004-6361/201935140}

\bibitem[{{Burgess} {et~al.}(2020){Burgess}, {B{\'e}gu{\'e}}, {Greiner},
  {Giannios}, {Bacelj}, \& {Berlato}}]{BBB19}
{Burgess}, J.~M., {B{\'e}gu{\'e}}, D., {Greiner}, J., {et~al.} 2020, Nature
  Astronomy, 4, 174, \dodoi{10.1038/s41550-019-0911-z}

\bibitem[{{Chand} {et~al.}(2020){Chand}, {Pal}, {Banerjee}, {Sharma}, {Tam}, \&
  {He}}]{CPB20}
{Chand}, V., {Pal}, P.~S., {Banerjee}, A., {et~al.} 2020, \apj, 903, 9,
  \dodoi{10.3847/1538-4357/abb5fc}

\bibitem[{{Chodorowski} {et~al.}(1992){Chodorowski}, {Zdziarski}, \&
  {Sikora}}]{CZS92}
{Chodorowski}, M.~J., {Zdziarski}, A.~A., \& {Sikora}, M. 1992, \apj, 400, 181,
  \dodoi{10.1086/171984}

\bibitem[{{Comisso} {et~al.}(2020){Comisso}, {Sobacchi}, \& {Sironi}}]{CSS20}
{Comisso}, L., {Sobacchi}, E., \& {Sironi}, L. 2020, \apjl, 895, L40,
  \dodoi{10.3847/2041-8213/ab93dc}

\bibitem[{{Crumley} {et~al.}(2019){Crumley}, {Caprioli}, {Markoff}, \&
  {Spitkovsky}}]{CCM19}
{Crumley}, P., {Caprioli}, D., {Markoff}, S., \& {Spitkovsky}, A. 2019, \mnras,
  485, 5105, \dodoi{10.1093/mnras/stz232}

\bibitem[{{Crumley} \& {Kumar}(2013)}]{CK13}
{Crumley}, P., \& {Kumar}, P. 2013, \mnras, 429, 3238,
  \dodoi{10.1093/mnras/sts581}

\bibitem[{{Daigne} {et~al.}(2011){Daigne}, {Bo{\v s}njak}, \& {Dubus}}]{DBG11}
{Daigne}, F., {Bo{\v s}njak}, {\v Z}., \& {Dubus}, G. 2011, \aap, 526, A110,
  \dodoi{10.1051/0004-6361/201015457}

\bibitem[{{Daigne} \& {Mochkovitch}(1998)}]{DM98}
{Daigne}, F., \& {Mochkovitch}, R. 1998, \mnras, 296, 275,
  \dodoi{10.1046/j.1365-8711.1998.01305.x}

\bibitem[{{Dereli-B{\'e}gu{\'e}} {et~al.}(2020){Dereli-B{\'e}gu{\'e}}, {Pe'er},
  \& {Ryde}}]{DPR20}
{Dereli-B{\'e}gu{\'e}}, H., {Pe'er}, A., \& {Ryde}, F. 2020, \apj, 897, 145,
  \dodoi{10.3847/1538-4357/ab9a2d}

\bibitem[{{Drenkhahn} \& {Spruit}(2002)}]{DS02}
{Drenkhahn}, G., \& {Spruit}, H.~C. 2002, \aap, 391, 1141,
  \dodoi{10.1051/0004-6361:20020839}

\bibitem[{{Florou} {et~al.}(2021){Florou}, {Petropoulou}, \&
  {Mastichiadis}}]{FPM21}
{Florou}, I., {Petropoulou}, M., \& {Mastichiadis}, A. 2021, \mnras,
  \dodoi{10.1093/mnras/stab1285}

\bibitem[{{Gasparyan} {et~al.}(2022){Gasparyan}, {B{\'e}gu{\'e}}, \&
  {Sahakyan}}]{GBS21}
{Gasparyan}, S., {B{\'e}gu{\'e}}, D., \& {Sahakyan}, N. 2022, \mnras, 509,
  2102, \dodoi{10.1093/mnras/stab2688}

\bibitem[{{Ghisellini} {et~al.}(2020){Ghisellini}, {Ghirlanda}, {Oganesyan},
  {Ascenzi}, {Nava}, {Celotti}, {Salafia}, {Ravasio}, \& {Ronchi}}]{GGO19}
{Ghisellini}, G., {Ghirlanda}, G., {Oganesyan}, G., {et~al.} 2020, \aap, 636,
  A82, \dodoi{10.1051/0004-6361/201937244}

\bibitem[{{Giannios}(2006)}]{Gia06}
{Giannios}, D. 2006, \aap, 457, 763, \dodoi{10.1051/0004-6361:20065000}

\bibitem[{{Giannios} \& {Uzdensky}(2019)}]{GU19}
{Giannios}, D., \& {Uzdensky}, D.~A. 2019, \mnras, 484, 1378,
  \dodoi{10.1093/mnras/stz082}

\bibitem[{{Goodman}(1986)}]{Goo86}
{Goodman}, J. 1986, \apjl, 308, L47, \dodoi{10.1086/184741}

\bibitem[{{Guetta} {et~al.}(2011){Guetta}, {Pian}, \& {Waxman}}]{GPW11}
{Guetta}, D., {Pian}, E., \& {Waxman}, E. 2011, \aap, 525, A53,
  \dodoi{10.1051/0004-6361/201014344}

\bibitem[{{Guiriec} {et~al.}(2015){Guiriec}, {Kouveliotou}, {Daigne}, {Zhang},
  {Hasco{\"e}t}, {Nemmen}, {Thompson}, {Bhat}, {Gehrels}, {Gonzalez}, {Kaneko},
  {McEnery}, {Mochkovitch}, {Racusin}, {Ryde}, {Sacahui}, \&
  {{\"U}nsal}}]{GKD15}
{Guiriec}, S., {Kouveliotou}, C., {Daigne}, F., {et~al.} 2015, \apj, 807, 148,
  \dodoi{10.1088/0004-637X/807/2/148}

\bibitem[{{Gupta} \& {Zhang}(2007)}]{GZ07}
{Gupta}, N., \& {Zhang}, B. 2007, \mnras, 380, 78,
  \dodoi{10.1111/j.1365-2966.2007.12051.x}

\bibitem[{{H{\"u}mmer} {et~al.}(2010){H{\"u}mmer}, {R{\"u}ger}, {Spanier}, \&
  {Winter}}]{HRS10}
{H{\"u}mmer}, S., {R{\"u}ger}, M., {Spanier}, F., \& {Winter}, W. 2010, \apj,
  721, 630, \dodoi{10.1088/0004-637X/721/1/630}

\bibitem[{{Kirk} {et~al.}(2000){Kirk}, {Guthmann}, {Gallant}, \&
  {Achterberg}}]{KGG00}
{Kirk}, J.~G., {Guthmann}, A.~W., {Gallant}, Y.~A., \& {Achterberg}, A. 2000,
  \apj, 542, 235, \dodoi{10.1086/309533}

\bibitem[{{Li}(2019)}]{Li19}
{Li}, L. 2019, \apjs, 242, 16, \dodoi{10.3847/1538-4365/ab1b78}

\bibitem[{{Lipari} {et~al.}(2007){Lipari}, {Lusignoli}, \& {Meloni}}]{LLM07}
{Lipari}, P., {Lusignoli}, M., \& {Meloni}, D. 2007, \prd, 75, 123005,
  \dodoi{10.1103/PhysRevD.75.123005}

\bibitem[{{Lundman} {et~al.}(2013){Lundman}, {Pe'er}, \& {Ryde}}]{LPR13}
{Lundman}, C., {Pe'er}, A., \& {Ryde}, F. 2013, \mnras, 428, 2430,
  \dodoi{10.1093/mnras/sts219}

\bibitem[{{Lyutikov} \& {Blackman}(2001)}]{LB01}
{Lyutikov}, M., \& {Blackman}, E.~G. 2001, \mnras, 321, 177,
  \dodoi{10.1046/j.1365-8711.2001.04190.x}

\bibitem[{{MAGIC Collaboration} {et~al.}(2019){MAGIC Collaboration}, {Acciari},
  {Ansoldi}, {Antonelli}, {Arbet Engels}, {Baack}, {Babi{\'c}}, {Banerjee},
  {Barres de Almeida}, {Barrio}, {Becerra Gonz{\'a}lez}, {Bednarek},
  {Bellizzi}, {Bernardini}, {Berti}, {Besenrieder}, {Bhattacharyya},
  {Bigongiari}, {Biland }, {Blanch}, {Bonnoli}, {Bo{\v s}njak}, {Busetto},
  {Carosi}, {Carosi}, {Ceribella}, {Chai}, {Chilingaryan}, {Cikota}, {Colak},
  {Colin}, {Colombo}, {Contreras}, {Cortina}, {Covino}, {D'Amico}, {D'Elia},
  {da Vela}, {Dazzi}, {de Angelis}, {de Lotto}, {Delfino}, {Delgado},
  {Depaoli}, {di Pierro}, {di Venere}, {Do Souto Espi{\~n}eira}, {Dominis
  Prester}, {Donini}, {Dorner}, {Doro}, {Elsaesser}, {Fallah Ramazani},
  {Fattorini}, {Fern{\'a}ndez-Barral}, {Ferrara}, {Fidalgo}, {Foffano},
  {Fonseca}, {Font}, {Fruck}, {Fukami}, {Gallozzi}, {Garc\'{\i}a L{\'o}pez},
  {Garczarczyk}, {Gasparyan}, {Gaug}, {Giglietto}, {Giordano}, {Godinovi{\'c}},
  {Green}, {Guberman}, {Hadasch}, {Hahn}, {Herrera}, {Hoang}, {Hrupec},
  {H{\"u}tten}, {Inada}, {Inoue}, {Ishio}, {Iwamura}, {Jouvin}, {Kerszberg},
  {Kubo}, {Kushida}, {Lamastra}, {Lelas}, {Leone}, {Lindfors}, {Lombardi},
  {Longo}, {L{\'o}pez}, {L{\'o}pez-Coto}, {L{\'o}pez-Oramas}, {Loporchio},
  {Machado de Oliveira Fraga}, {Maggio}, {Majumdar}, {Makariev}, {Mallamaci},
  {Maneva}, {Manganaro}, {Mannheim}, {Maraschi}, {Mariotti}, {Mart\'{\i}nez},
  {Masuda}, {Mazin}, {Mi{\'c}anovi{\'c}}, {Miceli}, {Minev}, {Miranda},
  {Mirzoyan}, {Molina}, {Moralejo}, {Morcuende}, {Moreno}, {Moretti},
  {Munar-Adrover}, {Neustroev}, {Nigro}, {Nilsson}, {Ninci}, {Nishijima},
  {Noda}, {Nogu{\'e}s}, {N{\"o}the}, {Nozaki}, {Paiano}, {Palacio},
  {Palatiello}, {Paneque}, {Paoletti}, {Paredes}, {Pe{\~n}il}, {Peresano},
  {Persic}, {Prada Moroni}, {Prand ini}, {Puljak}, {Rhode}, {Rib{\'o}}, {Rico},
  {Righi}, {Rugliancich}, {Saha}, {Sahakyan}, {Saito}, {Sakurai}, {Satalecka},
  {Schmidt}, {Schweizer}, {Sitarek}, {{\v S}nidari{\'c}}, {Sobczynska},
  {Somero}, {Stamerra}, {Strom}, {Strzys}, {Suda}, {Suri{\'c}}, {Takahashi},
  {Tavecchio}, {Temnikov}, {Terzi{\'c}}, {Teshima}, {Torres-Alb{\`a}}, {Tosti},
  {Tsujimoto}, {Vagelli}, {van Scherpenberg}, {Vanzo}, {Vazquez Acosta},
  {Vigorito}, {Vitale}, {Vovk}, {Will}, {Zari{\'c}}, \& {Nava}}]{MAA20}
{MAGIC Collaboration}, {Acciari}, V.~A., {Ansoldi}, S., {et~al.} 2019, \nat,
  575, 455, \dodoi{10.1038/s41586-019-1750-x}

\bibitem[{{Mannheim} \& {Schlickeiser}(1994)}]{MK94}
{Mannheim}, K., \& {Schlickeiser}, R. 1994, \aap, 286, 983.
\newblock \doarXiv{astro-ph/9402042}

\bibitem[{{Mastichiadis} {et~al.}(2005){Mastichiadis}, {Protheroe}, \&
  {Kirk}}]{MPK05}
{Mastichiadis}, A., {Protheroe}, R.~J., \& {Kirk}, J.~G. 2005, \aap, 433, 765,
  \dodoi{10.1051/0004-6361:20042161}

\bibitem[{{M{\'e}sz{\'a}ros} \& {Rees}(2000)}]{MR00}
{M{\'e}sz{\'a}ros}, P., \& {Rees}, M.~J. 2000, \apj, 530, 292,
  \dodoi{10.1086/308371}

\bibitem[{{M{\"u}cke} {et~al.}(2000){M{\"u}cke}, {Engel}, {Rachen},
  {Protheroe}, \& {Stanev}}]{MER00}
{M{\"u}cke}, A., {Engel}, R., {Rachen}, J.~P., {Protheroe}, R.~J., \& {Stanev},
  T. 2000, Computer Physics Communications, 124, 290,
  \dodoi{10.1016/S0010-4655(99)00446-4}

\bibitem[{{Nakar} {et~al.}(2009){Nakar}, {Ando}, \& {Sari}}]{NAS09}
{Nakar}, E., {Ando}, S., \& {Sari}, R. 2009, \apj, 703, 675,
  \dodoi{10.1088/0004-637X/703/1/675}

\bibitem[{{Narayan} \& {Kumar}(2009)}]{NK09}
{Narayan}, R., \& {Kumar}, P. 2009, \mnras, 394, L117,
  \dodoi{10.1111/j.1745-3933.2009.00624.x}

\bibitem[{{Oganesyan} {et~al.}(2018){Oganesyan}, {Nava}, {Ghirlanda}, \&
  {Celotti}}]{ONG18}
{Oganesyan}, G., {Nava}, L., {Ghirlanda}, G., \& {Celotti}, A. 2018, \aap, 616,
  A138, \dodoi{10.1051/0004-6361/201732172}

\bibitem[{{Oganesyan} {et~al.}(2019){Oganesyan}, {Nava}, {Ghirlanda},
  {Melandri}, \& {Celotti}}]{ONG19}
{Oganesyan}, G., {Nava}, L., {Ghirlanda}, G., {Melandri}, A., \& {Celotti}, A.
  2019, \aap, 628, A59, \dodoi{10.1051/0004-6361/201935766}

\bibitem[{{Paczynski}(1986)}]{Pac86}
{Paczynski}, B. 1986, \apjl, 308, L43, \dodoi{10.1086/184740}

\bibitem[{{Parsotan} \& {Lazzati}(2018)}]{PL18}
{Parsotan}, T., \& {Lazzati}, D. 2018, \apj, 853, 8,
  \dodoi{10.3847/1538-4357/aaa087}

\bibitem[{{Pe'er}(2015)}]{Pee15}
{Pe'er}, A. 2015, Advances in Astronomy, 2015, 907321,
  \dodoi{10.1155/2015/907321}

\bibitem[{{Pe'er} \& {Ryde}(2011)}]{PR11}
{Pe'er}, A., \& {Ryde}, F. 2011, \apj, 732, 49,
  \dodoi{10.1088/0004-637X/732/1/49}

\bibitem[{{Pe'er} \& {Waxman}(2005)}]{PW05}
{Pe'er}, A., \& {Waxman}, E. 2005, \apj, 628, 857, \dodoi{10.1086/431139}

\bibitem[{{Pe'Er} {et~al.}(2012){Pe'Er}, {Zhang}, {Ryde}, {McGlynn}, {Zhang},
  {Preece}, \& {Kouveliotou}}]{PZR12}
{Pe'Er}, A., {Zhang}, B.-B., {Ryde}, F., {et~al.} 2012, \mnras, 420, 468,
  \dodoi{10.1111/j.1365-2966.2011.20052.x}

\bibitem[{{Petropoulou}(2014)}]{Pet14}
{Petropoulou}, M. 2014, \mnras, 442, 3026, \dodoi{10.1093/mnras/stu1079}

\bibitem[{{Petropoulou} \& {Mastichiadis}(2015)}]{PM15b}
{Petropoulou}, M., \& {Mastichiadis}, A. 2015, \mnras, 447, 36,
  \dodoi{10.1093/mnras/stu2364}

\bibitem[{{Pitik} {et~al.}(2021){Pitik}, {Tamborra}, \& {Petropoulou}}]{PTP21}
{Pitik}, T., {Tamborra}, I., \& {Petropoulou}, M. 2021, \jcap, 2021, 034,
  \dodoi{10.1088/1475-7516/2021/05/034}

\bibitem[{{Preece} {et~al.}(1998){Preece}, {Briggs}, {Mallozzi}, {Pendleton},
  {Paciesas}, \& {Band}}]{PBM98}
{Preece}, R.~D., {Briggs}, M.~S., {Mallozzi}, R.~S., {et~al.} 1998, \apjl, 506,
  L23, \dodoi{10.1086/311644}

\bibitem[{{Razzaque} {et~al.}(2010){Razzaque}, {Dermer}, \& {Finke}}]{RDF10}
{Razzaque}, S., {Dermer}, C.~D., \& {Finke}, J.~D. 2010, The Open Astronomy
  Journal, 3, 150, \dodoi{10.2174/1874381101003010150}

\bibitem[{{Rees} \& {Meszaros}(1994)}]{RM94}
{Rees}, M.~J., \& {Meszaros}, P. 1994, \apjl, 430, L93, \dodoi{10.1086/187446}

\bibitem[{{Ryde} {et~al.}(2017){Ryde}, {Lundman}, \& {Acuner}}]{RLA17}
{Ryde}, F., {Lundman}, C., \& {Acuner}, Z. 2017, \mnras, 472, 1897,
  \dodoi{10.1093/mnras/stx2019}

\bibitem[{{Ryde} \& {Pe'er}(2009)}]{RP09}
{Ryde}, F., \& {Pe'er}, A. 2009, \apj, 702, 1211,
  \dodoi{10.1088/0004-637X/702/2/1211}

\bibitem[{{Ryde} {et~al.}(2010){Ryde}, {Axelsson}, {Zhang}, {McGlynn}, {Pe'er},
  {Lundman}, {Larsson}, {Battelino}, {Zhang}, {Bissaldi}, {Bregeon}, {Briggs},
  {Chiang}, {de Palma}, {Guiriec}, {Larsson}, {Longo}, {McBreen}, {Omodei},
  {Petrosian}, {Preece}, \& {van der Horst}}]{RAZ10}
{Ryde}, F., {Axelsson}, M., {Zhang}, B.~B., {et~al.} 2010, \apjl, 709, L172,
  \dodoi{10.1088/2041-8205/709/2/L172}

\bibitem[{{Sahu} \& {Fort\'{\i}n}(2020)}]{SF20}
{Sahu}, S., \& {Fort\'{\i}n}, C. E.~L. 2020, \apjl, 895, L41,
  \dodoi{10.3847/2041-8213/ab93da}

\bibitem[{{Samuelsson} {et~al.}(2021){Samuelsson}, {Lundman}, \&
  {Ryde}}]{SLR21}
{Samuelsson}, F., {Lundman}, C., \& {Ryde}, F. 2021, arXiv e-prints,
  arXiv:2111.01810.
\newblock \doarXiv{2111.01810}

\bibitem[{{Sari} {et~al.}(1996){Sari}, {Narayan}, \& {Piran}}]{SNP96}
{Sari}, R., {Narayan}, R., \& {Piran}, T. 1996, \apj, 473, 204,
  \dodoi{10.1086/178136}

\bibitem[{{Sari} {et~al.}(1998){Sari}, {Piran}, \& {Narayan}}]{SPN98}
{Sari}, R., {Piran}, T., \& {Narayan}, R. 1998, \apjl, 497, L17,
  \dodoi{10.1086/311269}

\bibitem[{{Sironi} {et~al.}(2013){Sironi}, {Spitkovsky}, \& {Arons}}]{SSA13}
{Sironi}, L., {Spitkovsky}, A., \& {Arons}, J. 2013, \apj, 771, 54,
  \dodoi{10.1088/0004-637X/771/1/54}

\bibitem[{{Totani}(1998)}]{Tot98}
{Totani}, T. 1998, \apjl, 509, L81, \dodoi{10.1086/311772}

\bibitem[{{Uhm} \& {Zhang}(2014)}]{UZ14b}
{Uhm}, Z.~L., \& {Zhang}, B. 2014, Nature Physics, 10, 351,
  \dodoi{10.1038/nphys2932}

\bibitem[{{Vianello} {et~al.}(2018){Vianello}, {Gill}, {Granot}, {Omodei},
  {Cohen-Tanugi}, \& {Longo}}]{VGG17}
{Vianello}, G., {Gill}, R., {Granot}, J., {et~al.} 2018, \apj, 864, 163,
  \dodoi{10.3847/1538-4357/aad6ea}

\bibitem[{{von Kienlin} {et~al.}(2020){von Kienlin}, {Meegan}, {Paciesas},
  {Bhat}, {Bissaldi}, {Briggs}, {Burns}, {Cleveland}, {Gibby}, {Giles},
  {Goldstein}, {Hamburg}, {Hui}, {Kocevski}, {Mailyan}, {Malacaria},
  {Poolakkil}, {Preece}, {Roberts}, {Veres}, \& {Wilson-Hodge}}]{vMP20}
{von Kienlin}, A., {Meegan}, C.~A., {Paciesas}, W.~S., {et~al.} 2020, \apj,
  893, 46, \dodoi{10.3847/1538-4357/ab7a18}

\bibitem[{{Vurm} \& {Beloborodov}(2016)}]{VB16}
{Vurm}, I., \& {Beloborodov}, A.~M. 2016, \apj, 831, 175,
  \dodoi{10.3847/0004-637X/831/2/175}

\bibitem[{{Waxman} \& {Bahcall}(1997)}]{WB97}
{Waxman}, E., \& {Bahcall}, J. 1997, \prl, 78, 2292,
  \dodoi{10.1103/PhysRevLett.78.2292}

\bibitem[{{Yu} {et~al.}(2015){Yu}, {van Eerten}, {Greiner}, {Sari}, {Narayana
  Bhat}, {von Kienlin}, {Paciesas}, \& {Preece}}]{YVG15}
{Yu}, H.-F., {van Eerten}, H.~J., {Greiner}, J., {et~al.} 2015, \aap, 583,
  A129, \dodoi{10.1051/0004-6361/201527015}

\bibitem[{{Zhang} \& {Yan}(2011)}]{ZY11}
{Zhang}, B., \& {Yan}, H. 2011, \apj, 726, 90,
  \dodoi{10.1088/0004-637X/726/2/90}

\bibitem[{{Zhang} \& {Zhang}(2014)}]{ZZ14}
{Zhang}, B., \& {Zhang}, B. 2014, \apj, 782, 92,
  \dodoi{10.1088/0004-637X/782/2/92}

\bibitem[{{Zhang} {et~al.}(2018){Zhang}, {Zhang}, {Castro-Tirado}, {Dai},
  {Tam}, {Wang}, {Hu}, {Karpov}, {Pozanenko}, {Zhang}, {Mazaeva}, {Minaev},
  {Volnova}, {Oates}, {Gao}, {Wu}, {Shao}, {Tang}, {Beskin}, {Biryukov},
  {Bondar}, {Ivanov}, {Katkova}, {Orekhova}, {Perkov}, {Sasyuk}, {Mankiewicz},
  {{\.Z}arnecki}, {Cwiek}, {Opiela}, {Zadro{\.Z}ny}, {Aptekar}, {Frederiks},
  {Svinkin}, {Kusakin}, {Inasaridze}, {Burhonov}, {Rumyantsev}, {Klunko},
  {Moskvitin}, {Fatkhullin}, {Sokolov}, {Valeev}, {Jeong}, {Park},
  {Caballero-Garc\'{\i}a}, {Cunniffe}, {Tello}, {Ferrero}, {Pandey},
  {Jel\'{\i}nek}, {Peng}, {S{\'a}nchez-Ramrez}, \& {Castell{\'o}n}}]{ZZC18}
{Zhang}, B.-B., {Zhang}, B., {Castro-Tirado}, A.~J., {et~al.} 2018, Nature
  Astronomy, 2, 69, \dodoi{10.1038/s41550-017-0309-8}

\end{thebibliography}

\end{document}